\documentclass[sigconf]{acmart}

\renewcommand\footnotetextcopyrightpermission[1]{} 
\usepackage{xspace}
\usepackage{colortbl} 
\usepackage{graphicx}
\usepackage{arydshln}     
\usepackage{makecell}
\usepackage{array}

\usepackage[english]{babel}

\usepackage{tikz}
\usepackage{tkz-euclide}
\usetikzlibrary{positioning,arrows.meta,calc}

\usepackage{float}
\usepackage{placeins}
\usepackage{subcaption}

\usepackage{pifont} 
\usepackage{bbding} 

\usepackage{url}

\newcommand{\figref}[1]{Fig.\ref{#1}\xspace}

\definecolor{theorycol}{RGB}{205,203,188}

\usepackage{comment}
\usepackage{filecontents}
\usepackage{multirow}

\usepackage{amsmath,amssymb,amsfonts,amsthm}
\usepackage{algorithmic}
\usepackage[linesnumbered, ruled]{algorithm2e}

\usepackage{textcomp}
\usepackage{xcolor}
\usepackage{booktabs}
\usepackage{dashbox}
\usepackage{todonotes}
\usepackage{enumitem}   
\usetikzlibrary{shapes.callouts}
\usepackage{listings}
\usepackage{trace}

\usepackage{mdframed}

\theoremstyle{plain}

\theoremstyle{definition}

\usepackage{threeparttable} 

\newcommand{\heading}[1]{{\vspace{3pt}\noindent{\textbf{#1}}}}

\tikzset{%
    pics/sema/.style args={#1/#2/#3}{code={%
        \ifstrequal{#2}{0}{%
            \node[circle,minimum width=1mm,draw,fill=#1] {};
        }{%
            \tkzDefPoint(0,0){O}
            \tkzDrawSector[R,fill=#1](O,1mm)(90,90-#2)
            \tkzDrawSector[R,fill=#3](O,1mm)(90-#2,90-360)
    }
    }},
}

\usepackage[
	n,
	operators,
	advantage,
	sets,
	adversary,
	landau,
	probability,
	notions,	
	logic,
	ff,
	mm,
	primitives,
	events,
	complexity,
	asymptotics,
	keys]{cryptocode}

\newenvironment{packeditemize}{
	\begin{itemize}}{\end{itemize}}

\title[Are Unreachable Nodes Truly Safe? \textit{Fully} Eclipsing Monero's P2P Network]{Are Unreachable Nodes Truly Safe? \textit{Fully} Eclipsing \\ Monero's P2P Network!}

\author{Ruisheng Shi$^{1}$, Jiaqi Zeng$^{1}$, Lina Lan$^{1}$, Shihan Zhang$^{1}$, Bing Han$^{1}$, \\Xiapu Luo$^{2}$, Qishu Jin$^{3}$, Wenliang Du$^{3}$, Qin Wang$^{4}$}
\thanks{Accepted by \textcolor{violet}{ACM Conference on Computer and Communications Security (CCS) 2026}. 
}
\affiliation{
\smallskip
\textit{$^1$Beijing University of Posts and Telecommunications} $|$ \textit{$^3$ Zhejiang University} \\ \textit{$^2$ Hong Kong Polytechnic University}   $|$ \textit{$^4$ CSIRO} \country{}
}

\begin{document}

\begin{abstract}
Eclipse attacks isolate a blockchain node by monopolizing its network connections. Existing attacks on Monero (NDSS'25), Bitcoin (USENIX'15/21, S\&P'20) and Ethereum (WWW'26) implicitly assume that the adversary can establish inbound connections, thereby excluding a large and practically dominant class of nodes: \textit{unreachable nodes} operating behind NATs. Such nodes are widely believed to enjoy stronger networks. We challenge this assumption and show that unreachability does NOT imply the expected resilience!

We present the first eclipse attacks tailored to unreachable nodes in Monero's P2P network. Our attacks require no inbound access to the victim. Instead, they first poison the peerlist of reachable nodes, which subsequently act as propagation relays to contaminate unreachable nodes' whitelists. The adversary then exploits Monero's built-in outbound connection refresh logic to evict benign neighbors and eventually monopolize all outbound connections. We instantiate this strategy in two attacks: \textsc{Nyx}, which targets long-running unreachable nodes and achieves a complete and persistent eclipse through network-wide poisoning; and \textsc{Moros}, a stealthier attack that exploits the bootstrapping phase to rapidly eclipse newly joined unreachable nodes.

We ethically evaluate both attacks. \textsc{Nyx} is validated via large-scale simulations on a Monero network constructed using the SEED Emulator, while \textsc{Moros} is demonstrated on the Monero mainnet against controlled targets. Our results show that unreachable nodes can be reliably driven into stable, long-lived eclipse states. We also propose countermeasures.

\end{abstract}

\maketitle
\section{Introduction}
\label{sec-intro}




Monero is best known for its transaction-layer privacy (e.g., stealth addresses~\cite{van2013cryptonote,eip5564}, and RingCT~\cite{sun2017ringct,yuen2020ringct}). However, its security also depends largely on the integrity of its peer-to-peer (P2P) network. Monero's peerlist propagation and connection management adopt relatively weak assumptions, which can be exploited at the network layer~\cite{dotan2021survey,neudecker2018network}. A canonical example is eclipse attacks~\cite{heilman2015eclipse}, where an adversary monopolizes a target's connections to control its network view and communication, enabling follow-on attacks such as deanonymization~\cite{biryukov2014deanonymisation,fanti2017deanonymization,biryukov2019deanonymization}, network partitioning~\cite{saad2021syncattack,heo2023partitioning,saad2023three}, and selfish-mining advantages~\cite{nayak2016stubborn,ritz2018impact,kang2021understanding,li2025does}

Prior work studied eclipse attacks in major cryptocurrency systems, including Bitcoin~\cite{heilman2015eclipse,tran2020stealthier,tran2021routing} and Ethereum~\cite{shi2026eclipse,dahlke2018low,henningsen2019eclipsing}. However, these attacks confront two fundimental limitations. (i) attackers typically lack control over when the eclipse occurs, as takeover depends on victim restarts or connection rebuilding. (ii) many schemes rely on weakly protected peerlist population paths and become ineffective once networks deploy more proactive defenses that restrict peer injection. 
Shi \emph{et al.}~\cite{shi2025eclipse} presented the eclipse attack on Monero by combining peerlist population with a connection-reset technique based on double-spending conflicts~\cite{bojja2017dandelion,fanti2018dandelion++}. While effective, the attack requires continuous active interference to maintain isolation, increasing both cost and detectability.

More fundamentally, existing eclipse attacks rely on a critical physical assumption: the adversary must be able to \textbf{establish direct incoming connections to the target node}. This constraint restricts prior schemes to \emph{reachable} nodes with public IP addresses, rendering them inapplicable to \emph{unreachable} nodes operating behind Network Address Translation (NAT) or firewalls~\cite{srisuresh1999ip}.

Unreachable nodes are commonly believed to present a reduced attack surface, as they cannot accept unsolicited inbound connections. In practice, a substantial fraction of Monero users operate nodes in private home or corporate networks~\cite{donet2014bitcoin,miller2015discovering,mariem2020all}, forming a \textit{silent majority} that remains largely invisible to external measurement tools~\cite{wang2017towards,grundmann2021estimating}.Our own mainnet measurement confirms their prevalence: $80.64\%$ of observed nodes are unreachable (Appendix~\ref{app:unreach}). This has led to the widespread assumption that unreachable nodes are inherently more secure.

We argue that this assumption is fundamentally flawed! Since unreachable nodes cannot accept incoming connections, they depend \textbf{entirely} on their outgoing peers for peerlist discovery and connection selection. Once these outgoing connections are compromised, unreachable nodes lack any protocol-level recovery path: honest external nodes cannot rediscover or reconnect with the victim. As a result, unreachable nodes face a \emph{self-healing dilemma}, in which isolation becomes stable and long-lasting once established.

To address this gap, we systematically analyze Monero's peerlist propagation and connection management mechanisms and present the first eclipse attack specifically targeting \textit{unreachable} nodes. Unlike prior approaches \cite{heilman2015eclipse,henningsen2019eclipsing,wust2016ethereum,shi2025eclipse,shi2026eclipse} requiring direct interaction with the target , our attack employs \emph{indirect peerlist poisoning}: the adversary \textbf{continuously injects malicious peer records into \emph{reachable} nodes} across the network, \textbf{turning them into propagation proxies that contaminate unreachable nodes} during routine P2P interactions.

For connection takeover, rather than relying on node restarts or adversarial resets, our attack exploits Monero's protocol-internal outbound connection refresh mechanism. Under this mechanism, nodes periodically refresh outbound connections by proactively dropping existing peers and selecting new ones from the peerlist. Since this process is triggered by normal protocol operation rather than explicit adversarial reset actions, our attack progresses steadily and achieves stable isolation over time.

We abstract the attack into three phases: (i) \emph{network-wide poisoning} of reachable nodes' peerlists; (ii) \emph{indirect contamination} of unreachable nodes via peerlist propagation; and (iii) \emph{connection reset}, which leverages the victim's connection churn logic to replace benign peers with malicious ones. 

This model covers two representative scenarios in the node lifecycle: (1) for \textit{existing} unreachable nodes with stable connectivity, the attack achieves gradual takeover through long-term peerlist infiltration; (2) for \textit{newborn} unreachable nodes, the attack enables immediate isolation during the bootstrapping phase, as their initial peerlists rely entirely on poisoned seed nodes. Due to the inherent inability of unreachable nodes to accept incoming connections, once their peerlists are fully compromised, they lose all paths to independently rediscover the honest network, falling into a stable and irreversible eclipse state. We accordingly instantiate the two attack variants, termed \textsc{Nyx} and \textsc{Moros}\footnote{%
\textsc{Nyx} (the Greek goddess of night) denotes network-wide poisoning that plunges unreachable nodes into global isolation, while \textsc{Moros} (the embodiment of doom) captures eclipse-at-birth via poisoned seed nodes.
}.

We evaluate attacks through large-scale emulation on a 1{,}200-node Monero network calibrated against mainnet latency, jitter, and loss measurements to study existing unreachable nodes, and through controlled experiments on the Monero mainnet to validate eclipse-at-birth against newborn unreachable nodes. Our results show that unreachable nodes can be reliably and persistently eclipsed under realistic conditions.

\vspace{3pt}
\begin{center}
 \fbox{
\begin{minipage}{0.9\linewidth}
\noindent\textbf{Responsible disclosure:} We
have reported our attacks and detected vulnerabilities to Monero
official team via Hackerone with the identifier \#3702913.

   \end{minipage}
}
\end{center}
\vspace{3pt}

In short, our main contributions are as follows:
\begin{packeditemize}
    \item We uncover structural vulnerabilities in Monero's peerlist propagation and connection management that disproportionately affect unreachable nodes.
    \item We present the first \emph{complete} eclipse attack against unreachable Monero nodes, achieving \emph{full takeover of all outgoing connections}. The attack combines indirect peerlist poisoning with protocol-driven connection rotation, enabling stable and sustained isolation without requiring incoming connections to the victim. Our attack is applicable towards a broader class of eclipse vulnerabilities affecting outbound-only nodes in modern P2P networks.
    \item We demonstrate that unreachable nodes can be \emph{fully} eclipsed throughout their entire lifecycle, evaluating attacks against both \emph{existing} unreachable nodes via large-scale emulation and \emph{newborn} unreachable nodes via controlled Monero mainnet experiments. In our experiments, \textsc{Nyx} reaches full takeover within \mbox{27}\,min, while \textsc{Moros} eclipses newborn unreachable nodes within 8\,s.
    \item We discuss the protocol mechanisms enabling the attack and propose corresponding mitigation strategies.
\end{packeditemize}

\section{Technical Warmups}
\label{sec-background-p2p}

We first review the Monero P2P mechanisms relevant to our attacks. We then explain how unreachable nodes change the attack surface.

\subsection{Monero's P2P Network (Sketch)}

We sketch key concepts; full details appear in Appendix~\ref{apdx-background-p2p}.

\heading{Unreachable nodes.} Monero nodes behind NATs or firewalls cannot accept incoming connections; they only establish outgoing connections to reachable peers (i.e., peers accepting inbound connections). Such a node establishes outgoing connections exclusively from local peerlist and depends entirely on outgoing peers for peer discovery. Once all outgoing peers are attacker-controlled, no honest node reaches the target, making eclipse states \textit{self-reinforcing}.

\heading{Peerlist management.} Each node maintains a \emph{whitelist} (up to 1{,}000 verified peers) and a \emph{graylist} (up to 5{,}000 unverified candidates). A peer enters the whitelist through three paths: (i)~incoming-connection validation (the receiver sends a PING; a valid PONG triggers insertion), (ii)~periodic graylist probing (\path|gray_peerlist_housekeeping()| randomly samples one graylist entry per minute and promotes it upon a successful handshake), and (iii)~outgoing-connection establishment. Whitelist entries are sorted by \texttt{last\_seen}; timed sync updates \texttt{last\_seen} for outgoing peers every 60\,s but \emph{not} for incoming peers, so current outgoing connections resist eviction while injected incoming-peer entries gradually age out. The graylist accepts records from handshake and timed sync responses without reachability verification and evicts the oldest entries first (FIFO) when capacity is reached. An attacker who controls an outgoing connection to the target can therefore inject up to 250 records per timed sync round, steadily displacing historical benign entries.

\heading{Peerlist propagation.} Handshake and timed sync responses each carry up to 250 peer records sampled from the sender's whitelist. For unreachable nodes, this process is strictly indirect: all peer information is learned from upstream reachable peers. Each connection maintains a \emph{sent-address list} that excludes previously transmitted records from subsequent responses. After sufficiently many rounds the sender's whitelist is exhausted, bounding the total benign-record input from any single honest outgoing peer.

\heading{Connection refresh.} Outgoing connections in Monero are not static. The protocol periodically refreshes outgoing connections via \path|update_sync_search()|, releasing one outgoing slot approximately every 101\,s and selecting a replacement from the local peerlist as part of normal operation, rather than in response to explicit failures or restarts (trigger details in Appendix~\ref{apdx-background-p2p}). The post-release count of~11 exceeds whitelist-priority threshold (default:~8), so the replacement always follows the \emph{graylist-first} path.

\heading{Seed nodes and bootstrapping.} A newborn node whose whitelist and graylist are both empty sends a handshake to a hardcoded \emph{seed node}\footnote{Monero hardcodes a small set of reachable-node addresses as seed nodes. They serve as initial peer-information sources during node startup and are used only when the local peerlist is empty or repeated connection attempts fail.} and writes the returned records (up to~250) into its graylist, which is the only source of candidates for its 12 initial outgoing connections. If the seed node's whitelist has been poisoned, the newborn node's entire candidate set consists of attacker-controlled records, enabling eclipse at birth.

\subsection{What Do Unreachable Nodes Change?}
\label{subsec-unreachable-challenge}

Unreachable nodes remove the direct inbound path used by prior Monero eclipse attacks, but they also make peer discovery more dependent on the node's current outbound peers. The attacker can no longer write records directly into the victim's peerlist. Malicious records must instead be planted in reachable nodes and then relayed to the target through ordinary timed sync responses.

This creates three challenges. First, peer injection must become transitive. Since NAT blocks incoming connections to the target, malicious records must be planted in reachable nodes and then carried to the unreachable target through normal timed sync responses. Second, network-wide filling must survive graylist deduplication. If all attacker records reuse the same port, they collapse into a small set of $\langle\text{IP},\text{port}\rangle$ entries; port diversity is therefore needed to keep the injected records distinct. Third, connection takeover cannot rely on active resets. The attacker must instead exploit Monero's ordinary outgoing-connection refresh logic, so that benign peers are replaced through normal protocol execution.

The bootstrapping case has the same structure. A newborn unreachable node derives its initial graylist from seed-node responses and selects its first outgoing peers from that list. Poisoned seed responses can therefore give the node an attacker-dominated peer view before it has any independent view of the network.

This is where our attacks enter. Instead of trying to contact the unreachable target directly, \textsc{Nyx} first compromises the peer-information sources around it: reachable nodes that may later serve as its outgoing peers. Their poisoned whitelists are then carried into the target through timed sync, after which Monero's own connection-refresh logic gradually replaces benign outgoing peers. \textsc{Moros} applies the same principle earlier in the lifecycle, entering through the seed-node bootstrap path before the newborn node has formed any independent peer view.

\section{Our Two Eclipse Attacks}
\label{sec-attack}

We present two eclipse attacks against Monero unreachable nodes: \textsc{Nyx}~(\S\ref{sec-nyx}) for existing nodes and \textsc{Moros}~(\S\ref{sec-moros}) for newborn nodes.

\subsection{Threat Model}
\label{sec-threat}

Our eclipse attack targets Monero \textit{unreachable nodes} behind a NAT or firewall. Such nodes can only initiate outgoing connections and cannot accept incoming connections from external peers. We categorize targets into two types based on operational state.

\begin{packeditemize}
    \item \textit{Newborn nodes.} Nodes that have just joined the Monero network with an empty peerlist.
    \item \textit{Existing nodes.} Nodes that have been operating for a period with 12 stable outgoing connections and a populated peerlist.
\end{packeditemize}

Since unreachable nodes cannot be actively connected via incoming requests, the attacker's objective is to occupy all 12 outgoing connections of the target node, thereby isolating it from the honest network~(\S\ref{sec-background-p2p}). The attacker controls 1{,}000 distinct /24 subnets (one IP per subnet) for whitelist filling, and 20 additional IPs for graylist filling. An asynchronous probing infrastructure (Appendix~\ref{app:probing}) maintains an inventory of approximately 3{,}000 reachable Monero nodes and supports the per-node port assignment in \textit{N-I}~(\S\ref{sec-n1}). The attacker does not need to know the target's IP address or establish any connection to it; \textit{N-I} poisons all reachable nodes uniformly, so a single deployment applies to arbitrary unreachable nodes.

\subsection{\textsc{Nyx} Attack}
\label{sec-nyx}

\begin{figure*}[t] 
    \centering
    \includegraphics[width=0.95\textwidth]{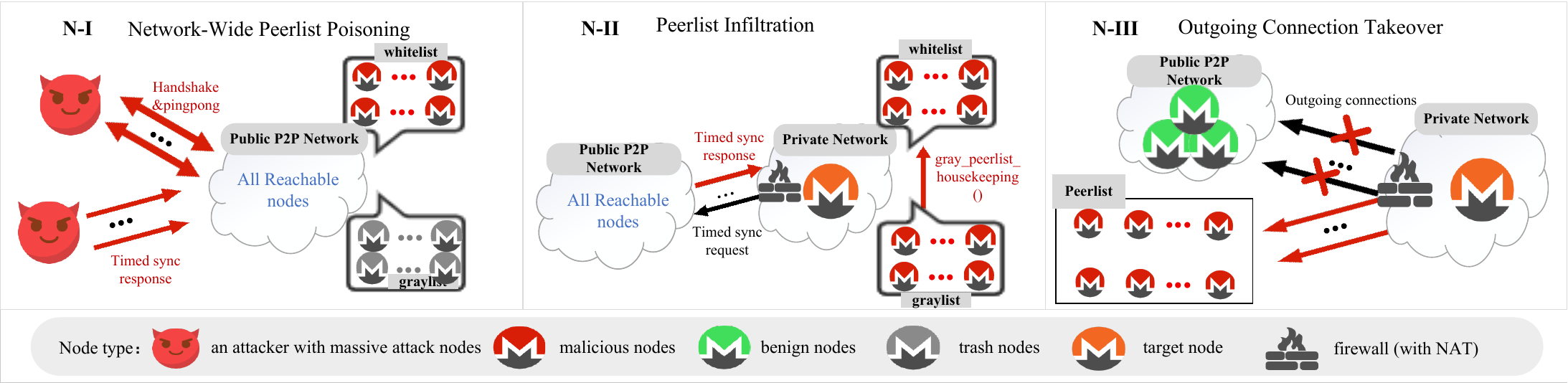}
    \caption{Overview of the \textsc{Nyx} attack.}
    \label{fig:nyx_overview}
\end{figure*}

\textsc{Nyx} targets existing unreachable nodes with stable outgoing connections. It composes the three shifts identified in \S\ref{sec-attack} into a three-phase pipeline (Fig.~\ref{fig:nyx_overview}):

\begin{packeditemize}
    \item[\ding{192}] \textbf{\textit{N-I}: network-wide peerlist poisoning.} It poisons reachable nodes' whitelists across the network through port-diverse filling, realizing insights (i) and (ii).
    \item[\ding{193}] \textbf{\textit{N-II}: peerlist infiltration.} It lets the protocol's timed sync mechanism carry the poisoning into the target's graylist, realizing insight (i) at the NAT boundary.
    \item[\ding{194}] \textbf{\textit{N-III}: outgoing connection takeover.} It invokes Monero's autonomous \path|update_sync_search()| to replace benign outgoing peers with malicious ones, realizing insight (iii).
\end{packeditemize}

Phase~\textit{N-I} must run first and continue running throughout the attack. Once \textit{N-I} has poisoned reachable nodes' whitelists, \textit{N-II} and \textit{N-III} proceed without further attacker intervention, driven by the target's own timed sync and outgoing-connection refresh.

\subsubsection{\underline{\textit{N-I}: network-wide peerlist poisoning}}
\label{sec-n1}
\textit{N-I} poisons reachable nodes' whitelists so that they relay attacker-controlled records to the target during timed sync.
It must satisfy three requirements: \emph{coverage}, reaching as many reachable nodes as possible; \emph{occupation}, dominating each poisoned whitelist with attacker records; and \emph{persistence}, continuously counteracting benign record propagation.
The attacker runs whitelist and graylist filling in parallel, with whitelist filling using a \emph{port-diversity} design.

\heading{\textit{N-I}-1: Whitelist poisoning against reachable nodes.}
The attacker initiates handshakes from 1{,}000 distinct IPs to
each reachable node; each IP passes \texttt{PING}/\texttt{PONG}
validation and is inserted into the receiver's whitelist
(\S\ref{sec-background-p2p}), following the primitive
of~\cite{shi2025eclipse}.
Since the whitelist capacity is 1{,}000 entries, a single pass
saturates one node's whitelist.
Scaling from a single target to the entire network introduces two
problems: the graylist's per-$\langle$IP, port$\rangle$
deduplication renders the filling ineffective, and filling
efficiency is too low.

\heading{Port-diversity design.}
Attacker records written into reachable nodes' whitelists must
pass the graylist's per-$\langle$IP, port$\rangle$ deduplication
to reach the target's graylist (\S\ref{sec-background-p2p}).
The port announced by each attacker IP during filling therefore
controls how many distinct records reach the target.

Directly reusing the single-target scheme of~\cite{shi2025eclipse}
across the network renders the filling ineffective.
Every attacker IP announces a fixed port~$p_0$ to every reachable
node, so all whitelists contain the same 1{,}000 records.
The target's 12 outgoing connections return up to $12 \times 250 =
3{,}000$ records per timed sync round, but after deduplication at
most 1{,}000 enter the graylist on round~1, and the injection rate
drops to zero from round~2 onward.

Port-diversity assigns each reachable node a dedicated announced
port. When filling that node, each attacker IP announces this dedicated
port. The same attacker IP therefore appears as a distinct
$\langle$IP, port$\rangle$ entry across different reachable nodes'
whitelists.

The target's 12 outgoing connections therefore return up to $12\times250=3{,}000$ mutually distinct $\langle\text{IP},\text{port}\rangle$ records per round, all of which pass graylist deduplication.

\heading{Concurrent filling and completeness guarantee.}
The attacker must complete filling before natural record
replacement dilutes the effect.
\texttt{PING}/\texttt{PONG} validation is asynchronous
(\S\ref{sec-background-p2p}). The attacker simultaneously initiates
handshakes from all 1{,}000 IPs to a given node, and the node
processes each \texttt{PING}/\texttt{PONG} independently without
blocking.
Both intra-node and inter-node filling execute in parallel, so
network-wide filling time is bounded by the attacker's concurrency
capacity.

Under high concurrency, TCP timeouts may cause some handshake or \texttt{PING} requests to fail, so one filling round may not write all 1{,}000 attacker IPs into a node's whitelist.
The attacker recovers missing entries by opening a fresh P2P connection and issuing consecutive timed sync requests.
Because \texttt{sent\_addresses} prevents duplicate records on the same connection (\S\ref{sec-background-p2p}), these responses progressively enumerate the node's whitelist.
The attacker compares the returned records with its 1{,}000 IPs and re-issues handshakes only for missing entries, repeating until all IPs are confirmed written.
Batch parameters and timeout settings are given in~\S\ref{sec-eavlua}.

\heading{\textit{N-I}-2: Graylist poisoning against reachable nodes.}
Whitelist filling alone cannot sustain the poisoning, because
\path|gray_peerlist_housekeeping()| continuously promotes graylist
records into the whitelist (\S\ref{sec-background-p2p}).
The attacker suppresses this channel by flooding every reachable
node's graylist with \emph{trash} records: randomly generated,
non-connectable $\langle$IP, port$\rangle$ entries.
The graylist filling follows the method of~\cite{shi2025eclipse}:
twenty additional attacker nodes maintain incoming connections and
return 250 trash records per timed sync response, injecting
$20 \times 250 = 5{,}000$ entries per round---sufficient to
saturate the graylist in a single pass. 
When \path|gray_peerlist_housekeeping()| samples a trash record,
the handshake fails because the address is non-connectable, wasting
the promotion slot.

\heading{Filling upper bound.}
The attacker cannot occupy all 1{,}000 whitelist slots.
Each reachable node keeps 12 stable outgoing-connection records whose \texttt{last\_seen} values are refreshed by timed sync every 60\,s (\S\ref{sec-background-p2p}).
Even if a filling round temporarily evicts them, the next refresh re-inserts them at the whitelist head.
Thus, the attacker's 1{,}000 IPs can occupy at most $1{,}000-12=988$ slots, giving
\begin{equation}
\label{eq:or-max}
\mathrm{OR}_{\max}=\frac{1{,}000-12}{1{,}000}\approx98.8\%.
\end{equation}
The actual OR may fall short of this bound when the attacker's reachable-node inventory is incomplete. Uncovered nodes propagate benign records from their own peerlists to poisoned nodes during routine timed sync, gradually diluting the occupation rate. Sustaining the bound requires continuous re-filling.

After \emph{N-I}, each poisoned reachable node's timed sync
responses deliver malicious records to the target at the
$\mathrm{OR}_{\max} \approx 98.8\%$ rate, providing the upstream
input for \emph{N-II}.
The port-diversity design also leaves the attacker holding a
malicious peer pool of approximately
$1{,}000 \times 3{,}000 = 3{,}000{,}000$ distinct
$\langle$IP, port$\rangle$ pairs---roughly $600\times$ the
graylist's 5{,}000-slot capacity.
This port count need not scale with the network: as we show in Appendix~\ref{app:sixgroup}, a fixed set of six ports suffices, bounding the adversarial budget at a constant $6{,}000$ node-identities regardless of network size.
\subsubsection{\underline{\textit{N-II}: peerlist infiltration}}
\label{sec-n2}
Once \emph{N-I} reaches steady state, the target's outgoing peers
deliver attacker-controlled records through routine timed sync
responses.
The goal of \emph{N-II} is to raise the malicious fraction of the
target's graylist to at least $\mathrm{IR}_{\min}$
(Eq.~\ref{eq:ir-min}) in the conservative case.

\heading{\textit{N-II}-1: Graylist infiltration.}
The target sends timed sync requests to its 12 outgoing peers every 60\,s.
Each peer returns up to 250 whitelist records, and the target writes unseen $\langle\text{IP},\text{port}\rangle$ entries into its graylist (\S\ref{sec-background-p2p}).
With port diversity, the 12 outgoing peers return mutually distinct attacker records, so these records pass graylist deduplication and are all inserted.

\textit{Quantitative analysis.}
Each timed sync round injects up to $12 \times 250 = 3{,}000$ new
records into the target's 5{,}000-slot graylist, which follows FIFO
eviction (\S\ref{sec-background-p2p}).
Under $\mathrm{OR}_{\max} \approx 98.8\%$, each reachable node's
whitelist retains approximately 12 benign records corresponding to
its own stable outgoing peers.
Each timed sync response samples 250 from 1{,}000 whitelist entries,
returning approximately 3 benign records in expectation.

In the optimistic case, the 12 outgoing connections return no benign
record across the first two rounds; approximately two minutes then
suffice to inject 6{,}000 malicious records, exceeding graylist
capacity and evicting all benign entries.
In the conservative case, the \texttt{sent\_addresses} mechanism
(\S\ref{sec-background-p2p}) drains each peer's 1{,}000-entry
whitelist in approximately four minutes, during which its 12 benign
records are necessarily returned.
The target's 12 outgoing connections collectively deliver at most
$12^2 = 144$ distinct benign records over the complete drain.
Since benign records arrive gradually across multiple rounds, FIFO
eviction driven by concurrent malicious injections may displace some
early-arriving benign entries before the drain completes; 144
therefore represents the upper bound on benign records present in the
graylist at any point.
We define this as the benign-record upper bound:
\begin{equation}
\label{eq:b-max}
B_{\max} = N_{\text{out}}^2 = 12^2 = 144.
\end{equation}
The corresponding malicious graylist fraction is:
\begin{equation}\label{eq:ir-min}
\mathrm{IR}_{\min} = \frac{5{,}000 - B_{\max}}{5{,}000}
= \frac{5{,}000 - 144}{5{,}000} \approx 97.1\%.
\end{equation}

\heading{\textit{N-II}-2: Whitelist infiltration.}
The target does not accept incoming connections, so the attacker
cannot write records into its whitelist directly.
Malicious records enter the whitelist only through the
Gray$\to$White promotion paths (\S\ref{sec-background-p2p}).
Once the graylist malicious fraction reaches $\mathrm{IR}_{\min}$,
both paths select attacker-controlled entries at the corresponding
rate, gradually injecting malicious records into the whitelist.

After \emph{N-II}, the target's graylist contains at most
$B_{\max} = 144$ benign records (Eq.~\ref{eq:b-max}), yielding a
malicious fraction of at least $\mathrm{IR}_{\min} \approx 97.1\%$
(Eq.~\ref{eq:ir-min}).
A subset of these malicious entries has already reached the whitelist
through Gray$\to$White promotion.

\subsubsection{\underline{\textit{N-III}: outgoing connection takeover}}
\label{subsec-nyx-phase3}

\textit{N-III} has two goals: to take over all 12 outgoing connections of the target, and to use malicious connections to evict residual benign graylist records, further infiltrate the whitelist, and make the eclipse states stable.

\heading{Drop-and-replace mechanism.} A synchronized node that holds 12 outgoing connections invokes \path|update_sync_search()| approximately every 101\,s, dropping one synchronized outgoing peer and selecting a replacement via the graylist-first path (\S\ref{sec-background-p2p}). The attacker does not need to actively tear down the target's benign connections, since the protocol performs the drop autonomously during normal operation. The attacker therefore needs every malicious outgoing peer to be classified as \texttt{state\_normal} on the target, so that \path|update_sync_search()| keeps replacing the remaining benign outgoing connections. To this end, each malicious node returns the genesis block as its top block in every handshake and timed-sync response. The target itself holds the genesis block, so it marks these peers as \texttt{state\_normal} on every protocol round, and the \path|update_sync_search()| trigger condition continues to hold throughout the attack. The core of \textit{N-III} is therefore to ensure that each replacement round is more likely to select an attacker-controlled node.

\begin{figure*}[t] 
    \centering
    \includegraphics[width=0.95\textwidth]{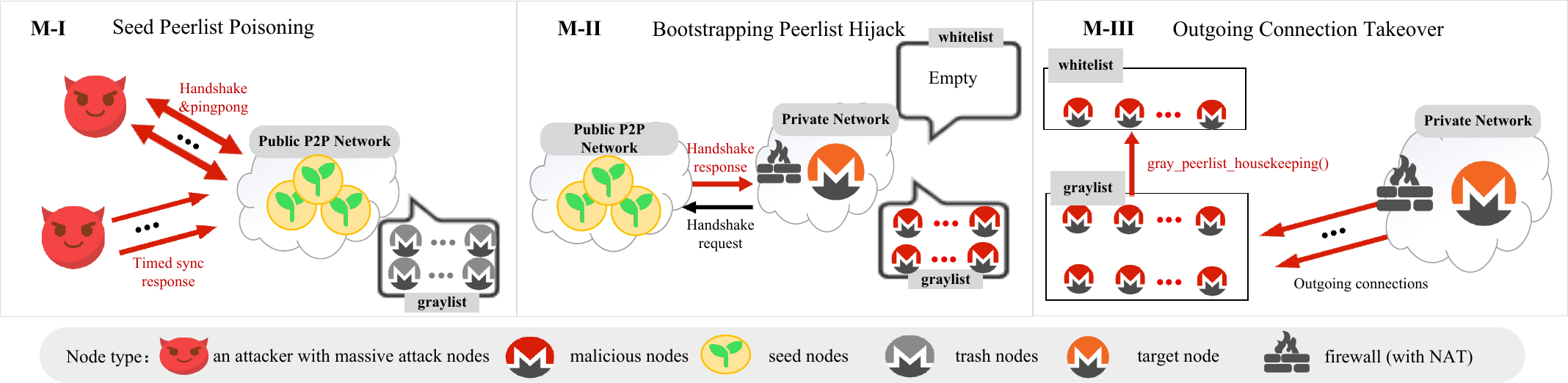}
    \caption{Overview of the \textsc{Moros} attack.}
    \label{fig:moros_overview}
\end{figure*}

\heading{Probability model.}
We model the drop-and-replace process as a probabilistic transition over $k$, the number of attacker-controlled outgoing connections. The model specifies the transition probability, the initial state produced by \textit{N-II}, and the condition under which $k$ increases in expectation.

Let $k$ ($0 \leq k \leq 12$) denote the number of attacker-controlled outgoing connections at the target. In each replacement round, the protocol applies per-IP and /24 subnet filtering to the graylist before sampling a replacement uniformly at random (see \S\ref{sec-background-p2p}). Let $M$ denote the number of malicious candidates remaining after filtering, and $B$ the corresponding number of benign candidates. The probability that a single round selects a malicious replacement is
\begin{equation}
\label{eq:pk}
P_k = \frac{M}{M + B}.
\end{equation}
and the probability that the dropped connection is itself malicious is $q_k = k/12$. The expected single-step change in $k$ is
\begin{equation}
\mathbb{E}[\Delta k] = (1 - q_k) P_k - q_k (1 - P_k) = P_k - q_k.
\label{eq:convergence}
\end{equation}

\heading{Initial state.}
\textit{N-II} leaves the target's graylist with at most $B_{\max}=144$ benign records (Eq.~\ref{eq:b-max}) and a malicious fraction of at least $\mathrm{IR}_{\min}$ (Eq.~\ref{eq:ir-min}), i.e., at least $4{,}856$ malicious records.
After /24 subnet filtering, the malicious candidates reduce to the attacker's $1{,}000$ IPs spanning $1{,}000$ distinct /24 subnets, so $M=1{,}000$.
The benign candidate count satisfies $B\le B_{\max}$ and is typically smaller, since benign records often cluster in fewer subnets than they have nodes.
With $M=1{,}000$ and $B\le144$, the initial replacement probability is at least $P_0\ge 1{,}000/(1{,}000+144)\approx87.4\%$.

\textit{Convergence condition.} With $M$ and $B$ instantiated, the model dictates that $k$ grows monotonically in expectation whenever $P_k > q_k$. Since $M$ is bounded by the attacker's IP count, raising $P_k$ requires shrinking $B$. The next part of the analysis shows that $B$ decreases over successive rounds.

\heading{Benign-candidate evolution.}
It remains to show that $B$ decreases over replacement rounds.
Appendix~\ref{app:niii-convergence} derives the round-level change $\Delta B$ from graylist FIFO eviction and the records returned by the newly selected peer.
If the replacement is malicious (Case A), it returns fresh malicious records from the \textit{N-I} peer pool, giving $\Delta B=-B/10$.
If the replacement is benign (Case B), it removes one benign graylist entry but may return a few benign records from its contaminated whitelist, giving the upper bound $\Delta B\le 5-B/10$.
Thus, Case A always reduces $B$, while Case B can increase $B$ only when $B<50$.
Because the initial malicious-replacement probability is at least $87.4\%$, Case A dominates early.
As $B$ shrinks, $P_k$ rises and the convergence margin $P_k-q_k$ widens.

\begin{figure}[t]
  \centering
  \includegraphics[width=\linewidth]{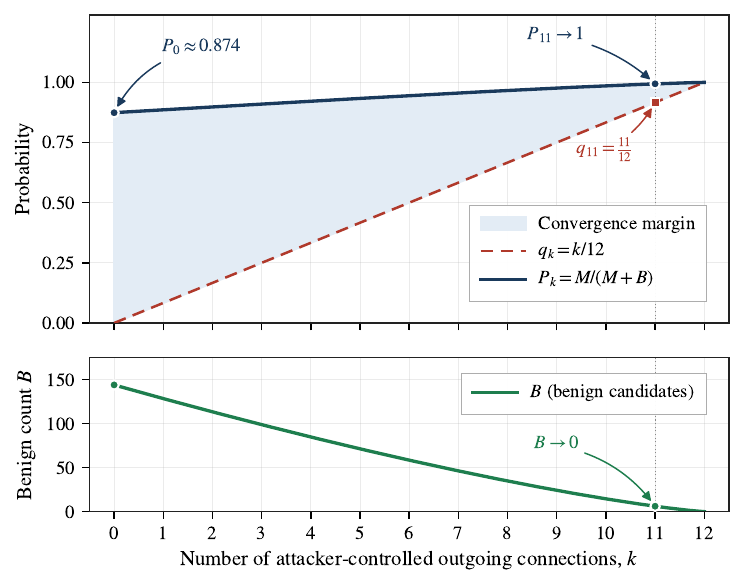}
  \caption{Expected evolution of $P_k$ (top) and $B$ (bottom) across the twelve replacement rounds of \textit{N-III}. Shaded region marks the margin $P_k - q_k$.}
  \label{fig:n3-convergence}
\end{figure}

\heading{Positive feedback.} Each newly established malicious outgoing connection samples up to 250 records from the malicious peer pool per timed sync cycle and returns them to the target, serving as a persistent malicious injection channel. As $k$ increases by one, one benign outgoing connection is lost, and one benign injection channel disappears with it. The two effects compound, so the rate at which benign records are evicted from the graylist grows with $k$, forming a positive feedback loop. Figure~\ref{fig:n3-convergence} visualizes this convergence across all values of $k$.

\textit{Boundary verification.} The most stringent case occurs at $k = 11$, where $q_{11} = 11/12 \approx 0.917$. The 11 malicious outgoing connections together inject $11 \times 250 = 2{,}750$ entirely new malicious records into the graylist per timed sync round. The sole remaining benign outgoing connection can contribute at most 12 distinct benign records over its entire lifetime, bounded by the benign entries in its whitelist (see \S\ref{sec-background-p2p}, \texttt{sent\_addresses} mechanism). Under FIFO eviction driven by the $2{,}750$-per-round malicious injection, $B$ approaches 0, so $P_{11} \to 1 > q_{11}$. Since $P_k$ rises monotonically with decreasing $B$ (\textit{Positive feedback}), the convergence condition $P_k > q_k$ holds at $k = 11$ and therefore at every $k \leq 11$ as well.

After \emph{N-III}, the target is eclipsed. All 12 outgoing connections are occupied by attacker-controlled peers, and its graylist and whitelist are dominated by malicious records. Since the target is unreachable and does not accept incoming connections, no benign node can establish a connection to break the eclipse.


\subsection{\textsc{Moros} Attack}
\label{sec-moros}

\textsc{Moros} targets newborn unreachable nodes and eclipses them at birth by hijacking the bootstrapping process.

A newborn node's entire initial graylist comes from a single seed-node handshake, which returns at most 250 records. Poisoning the small set of hardcoded seed nodes therefore suffices to dominate the newborn's peerlist before any honest peer interaction. Unlike \textsc{Nyx}, which propagates poisoning across the entire reachable-node population, \textsc{Moros} exploits this structural bottleneck to reach eclipse without touching the broader network.

\textsc{Moros} operates in three phases (Fig.~\ref{fig:moros_overview}):

\begin{packeditemize}
    \item[\ding{192}] \textit{\textbf{{M-I}: seed peerlist poisoning.}} The attacker poisons the hardcoded seed nodes' whitelists by applying \textit{N-I}'s filling procedure at a smaller scale.
    \item[\ding{193}] \textit{\textbf{M-II: bootstrapping-peerlist hijack.}} The newborn target receives a graylist dominated by attacker-controlled records during its initial seed handshake.
    \item[\ding{194}] \textit{\textbf{{M-III: outgoing connection takeover.}}} The attacker occupies the target's initial outgoing connections from the poisoned graylist and then uses \path|update_sync_search()| to replace any residual benign connection.
\end{packeditemize}

\textit{M-I} runs continuously throughout the attack, so that any newborn joining the network receives a poisoned seed response. Once \textit{M-I} is in steady state, \textit{M-II} and \textit{M-III} proceed without further attacker action, triggered by the newborn's own startup sequence.

\subsubsection{\underline{\textit{M-I}: seed peerlist poisoning}}
\label{sec-m1}

\textit{M-I} applies the whitelist-filling and graylist-filling procedures of \emph{N-I} (\S\ref{sec-n1}) to the six hardcoded Monero seed nodes. The filling targets are reduced from roughly $3{,}000$ reachable nodes to 6, so saturation is reached rapidly. The same 98.8\% occupation upper bound derived in \S\ref{sec-n1} applies, since seed nodes share the whitelist capacity and stable-outgoing structure of ordinary reachable nodes. A newborn node's subsequent seed handshake therefore returns a peerlist consisting almost entirely of attacker-controlled records.

\subsubsection{\underline{\textit{M-II}: bootstrapping-peerlist hijack}}
\label{sec-m2}

\textit{M-II} focuses on propagating malicious records from seed nodes to the target's local peerlist. Since a newborn node begins with an empty local whitelist and graylist, it must initiate a handshake with hardcoded seed nodes to retrieve an initial peerlist. 

Due to the persistent poisoning of the seed nodes, the peerlist response received by the target consists almost entirely of malicious records. Upon receiving these 250 initial records, the target node automatically writes them into its local graylist as per the protocol. These records then become the exclusive candidate pool for establishing the target's first 12 outgoing connections. This ``pollution transfer'' from the network infrastructure to the local list completes the indirect infiltration and sets the stage for final hijacking.

\subsubsection{\underline{\textit{M-III}: outgoing connection takeover}}
\label{sec-m3}
\emph{M-III} drives newborns from a cold start to a fully eclipsed state across three time phases: initial selection of the 12 outgoing connections from the poisoned graylist, persistence of the eclipse during initial block synchronization, and convergence to $k = 12$ once synchronization completes.

\heading{Initial outgoing selection.} At startup, the newborn node has an empty whitelist and anchorlist, so its initial outgoing connections can only be selected from the graylist written by \textit{M-II} (\S\ref{sec-background-p2p}). Since \textit{M-I} has already poisoned the seed response, the target's graylist is dominated by malicious records at startup, and the initial connection selection is highly likely to choose malicious peers. After a malicious outgoing connection is established, it continues to return records from the malicious peer pool in handshake and timed sync responses. These records keep entering the target's graylist and FIFO-evict residual benign records.

If all 12 initial outgoing connections are malicious, \textit{M-III} completes the eclipse immediately. If one benign outgoing connection remains after startup, i.e., $k=11$, the target enters the boundary state for the final takeover step. We show below that this state still converges to $k=12$. Thus, a small deviation in the initial selection affects only the completion time, not the takeover logic.

\heading{Persistence during initial sync.} If the target keeps one benign outgoing connection after startup ($k=11$), \path|update_sync_search()| does not actively replace it during initial synchronization, because the target's local block height remains below the network height (\S\ref{sec-background-p2p}). However, this benign connection has bounded influence: by the \texttt{sent\_addresses} mechanism, it can return at most 1{,}000 distinct benign records from its whitelist over its lifetime. In contrast, the 11 malicious outgoing connections inject up to $11\times250=2{,}750$ malicious records per timed sync round. Within two rounds, these injections exceed the 5{,}000-entry graylist capacity and evict residual benign records, so the benign graylist count approaches $B\approx0$ by the end of initial synchronization.

\heading{After synchronization.} After synchronization completes, \path|update_sync_search()| resumes its drop-and-replace cycle. At this point, the target is in the same state as the \textit{N-III} boundary case: $k=11$, with almost no benign candidates left in the graylist. Thus, the boundary condition in \S\ref{subsec-nyx-phase3} applies directly: $P_{11}\approx1>q_{11}=11/12$. The target's outgoing connections eventually converge to 12 malicious connections.

\section{Evaluation of Our Attack}
\label{sec-eavlua}
We experimentally evaluate both attacks. Since executing \textsc{Nyx} at scale would disrupt the operation of benign Monero nodes on the mainnet, we evaluate it on a large-scale Monero emulation platform we built. We evaluate \textsc{Moros} on the Monero mainnet under precautionary measures designed to avoid impacting the network.

\subsection{Experiment Setup}
\label{subsec-setup}

\subsubsection{\underline{Emulation environment}}
\label{subsubsec-sim-setup}

We build a controlled Monero P2P environment using the \textit{SEED Emulator}~\cite{github-seedemulator}, consisting of 100 ASes with BGP routing and 1{,}200 Monero nodes (including 6 seed nodes) running Monero v0.18.4.3~\cite{github-monero-v01843}.
Since all 1{,}200 containers run on a single physical server with sub-millisecond baseline latency, we inject per-node network impairments (delay, jitter, packet loss) calibrated against measurements from the Monero mainnet to approximate realistic propagation conditions (Appendix~\ref{app:netem}).

We set one node as the target. To simulate unreachable node behind NAT, we enforced network-layer access control via firewall rules to block all incoming connection requests. 
The attacker controls 1{,}000 distinct /24 subnets (one IP per subnet) for whitelist filling, and 20 additional IPs for graylist filling. Malicious nodes are implemented using \textit{py-levin}~\cite{github-pylevin}. The target's peerlists and outgoing connections are sampled every 30\,s via Monero RPC.

\subsubsection{\underline{Mainnet environment}}
\label{subsubsec-mainnet-setup}

We evaluate \textsc{Moros} against a controlled newborn target on the Monero mainnet.

The target runs on a public server with \texttt{--hide-my-port} enabled, which suppresses its P2P listening port to simulate unreachability.

The attacker uses 1{,}000 distinct public IPs to perform parallel whitelist poisoning against the six official Monero seed nodes. We do not perform graylist poisoning on the mainnet: a newborn node's initial graylist is bounded to the 250 records returned by a single seed handshake (\S\ref{sec-m2}), so dominating the seed-node whitelists is by itself sufficient for eclipse-at-birth. This design also confines the attack's footprint on the mainnet to the six seed nodes.

Since our 1{,}000 public IPs are drawn from a narrow set of /24 subnets, we disable the target's /24 subnet-diversity restriction during outbound connection selection. All other code paths follow the stock Monero client. Implementation details are in Appendix~\ref{app-impl-details}.

\subsection{Evaluation Metrics}
\label{subsec-metrics}

We quantify attack performance with the following metrics:
\begin{packeditemize}
    \item \textit{Peerlist occupation rate (OR).}
    The fraction of attacker-controlled peer records in a node's peerlist, measured separately for the 1{,}000-slot \textit{whitelist} and 5{,}000-slot \textit{graylist}.
    \item \textit{Connection takeover rate (CTR).}
    The fraction of attacker-controlled peers among the target's 12 outgoing peers. A CTR of 100\% (12/12) indicates a successful eclipse. We report the time at which CTR first reaches 12/12 as the Time-to-Eclipse (TTE).
    \item \textit{Eclipse stability.}
    The duration over which CTR = 100\% is maintained without a single benign outgoing connection reappearing.
\end{packeditemize}

\subsection{Evaluation of the \textsc{Nyx} Attack}
\label{subsec-eval-nyx}

\subsubsection{\underline{Evaluating \textit{N-I}}  (peerlist poisoning)}
\label{subsec-eval-nyx-phase1}
Before the attack starts, the states of reachable nodes and target node are in Table~\ref{tab:pre_attack_state}.

\begin{table}[!]
\centering
\caption{Pre-attack state.}
\label{tab:pre_attack_state}
\begin{tabular}{cccc}
\toprule
\textbf{Node type} & \textbf{Outgoing} & \textbf{Whitelist} & \textbf{Graylist} \\
\cmidrule{2-4}
Reachable nodes & 12 & 847/1000 & 353/5000 \\
Unreachable node & 12 & 483/1000 & 717/5000 \\
\bottomrule
\end{tabular}
\end{table}


\heading{Whitelist poisoning against reachable nodes.}
The attacker uses 1{,}000 distinct IPs to fill the whitelists of
the 1{,}199 reachable nodes; deployment details are in
App.~\ref{app-impl-details}. To pass the per-$\langle\text{IP},\text{port}\rangle$
deduplication on the target's graylist, we assign each reachable
node a distinct port: when filling node $r$, all 1{,}000 attacker
IPs use port $p_r$. The same attacker IP thereby produces
distinct $\langle\text{IP},\text{port}\rangle$ records under different nodes'
ports, all passing the target's deduplication, and the total
number of such records reaches approximately
$1{,}000 \times 1{,}199 \approx 1.2$M.

We process the 1{,}199 reachable nodes in batches of 120, with all
1{,}000 attacker IPs targeting the 120 nodes of each batch in
parallel and a 30\,s per-batch timeout. One complete filling round
(10 batches) takes approximately 5\,min, and rounds are launched
continuously to sustain the poisoning over time (\S\ref{sec-n1},
persistence). We monitored the attack for 18.2\,h, sampling every
reachable node's whitelist and graylist via RPC every 30\,s.

\begin{figure}[t]
    \centering
    \includegraphics[width=\linewidth]{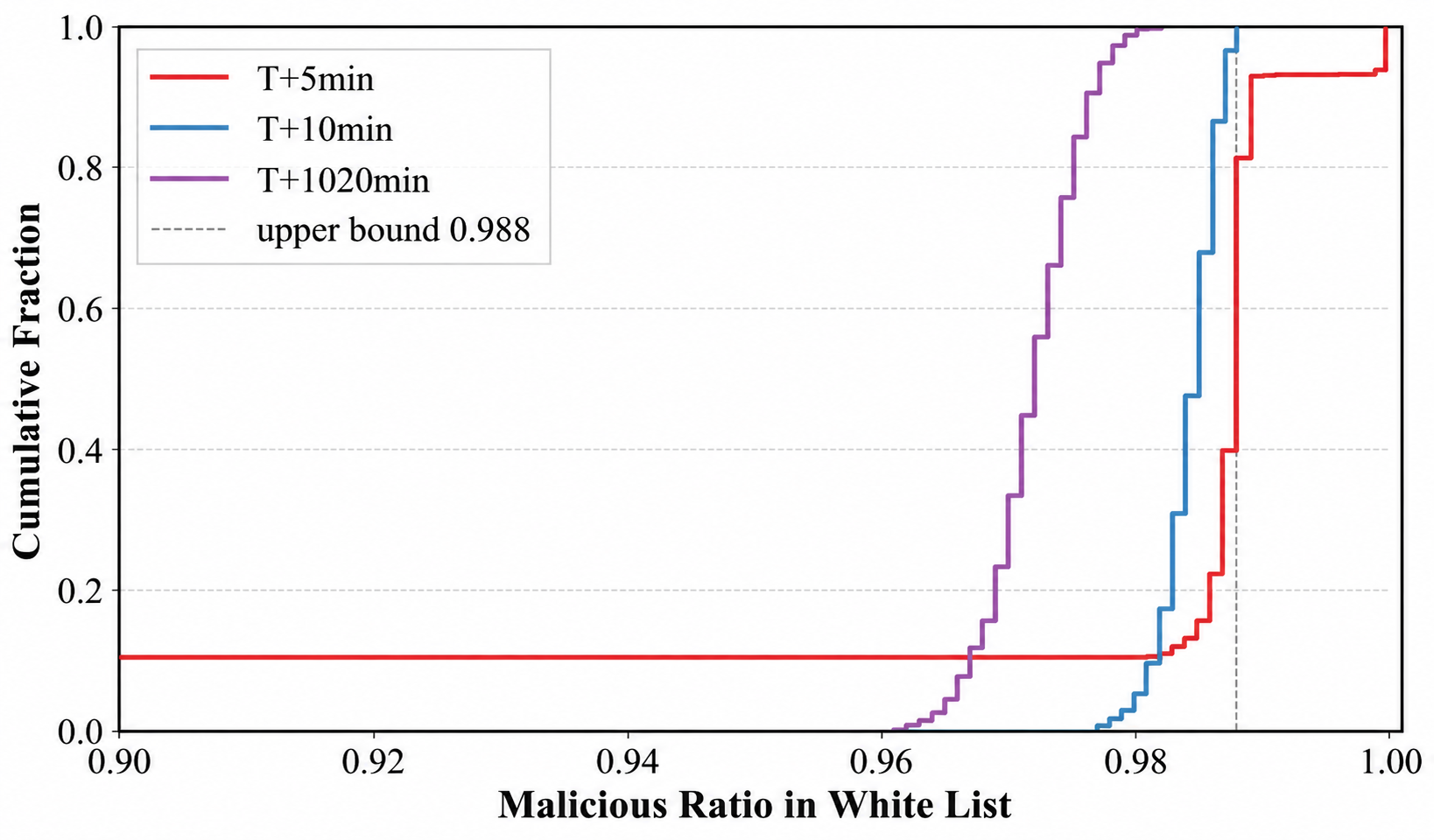}
    \caption{Evaluating \textit{N-I-1}: malicious-record occupation in reachable nodes' whitelists. Most nodes reach high occupation within 10min, approaching 98.8\% upper bound (Eq.~\ref{eq:or-max}).}
    \label{fig:nyx_reachable_white}
\end{figure}

\textit{Coverage.}
At T\,$+$5\,min, the first filling round has completed and 89.6\% of reachable nodes (1{,}074/1{,}199) have reached a malicious whitelist OR of at least 95\% (\figref{fig:nyx_reachable_white}).
The remaining 10.4\% received handshakes but did not reach full saturation because TCP timeouts during high-concurrency parallel filling caused a fraction of the 1{,}000 \texttt{PING} validations to fail.
The retry mechanism recovers these missing IPs in subsequent rounds.
By T\,$+$10\,min, 99.5\% of reachable nodes (1{,}193/1{,}199) reach an OR of at least 95\%, satisfying the \emph{coverage} requirement.

\textit{Occupation.}
The median OR across all reachable nodes reaches 98.5\% at T\,$+$10\,min (\figref{fig:nyx_reachable_white}), matching the 98.8\% theoretical upper bound derived in \S\ref{sec-n1}.
The residual 1.5\% benign fraction corresponds to the 12 stable outgoing connections per reachable node, whose \texttt{last\_seen} is continuously refreshed by timed sync and therefore resists eviction.
The attacker stably occupies the remaining 988/1{,}000 whitelist slots, satisfying the \emph{occupation} requirement.

\textit{Persistence.}
The CDF at T\,$+$1020\,min shifts leftward relative to T\,$+$10\,min (\figref{fig:nyx_reachable_white}).
The median OR declines from 98.5\% to 97.2\%, and the mean from 98.16\% to 96.90\%. This decline results from routine timed sync between reachable nodes, during which peers exchange peerlist records that include a small number of benign entries corresponding to their own outgoing connections.
These benign records gradually accumulate in reachable nodes' whitelists, displacing attacker entries.
The effect is bounded: even at T\,$+$1020\,min, 99.5\% of reachable nodes maintain an OR of at least 95\%.
The poisoned whitelists continue to deliver malicious records to the target at a rate sufficient for \textit{N-II}, satisfying the \emph{persistence} requirement.

\heading{Graylist poisoning against reachable nodes.}
We deploy 20 simulated nodes that maintain incoming connections to all reachable nodes and respond to every timed sync request with 250 randomly generated trash records. Across the 18.2\,h run, trash records occupy a mean of 98.15\% of each reachable node's graylist (malicious 0.67\%, benign 1.18\%), reaching 97.38\% within 1.5\,min and stabilizing at 99.02\% by minute 5. The residual benign fraction stems from routine timed sync between reachable nodes and from seed-node fallback when peerlist connections fail; these effects are bounded and do not undermine the suppression of \path|gray_peerlist_housekeeping()|. Per-node time series are available in our open-science repository.

\subsubsection{\underline{Evaluating \textit{N-II}} (peerlist infiltration)}
\label{subsec-eval-nyx-phase2}

\begin{figure}[t]
    \centering
    \includegraphics[width=\linewidth]{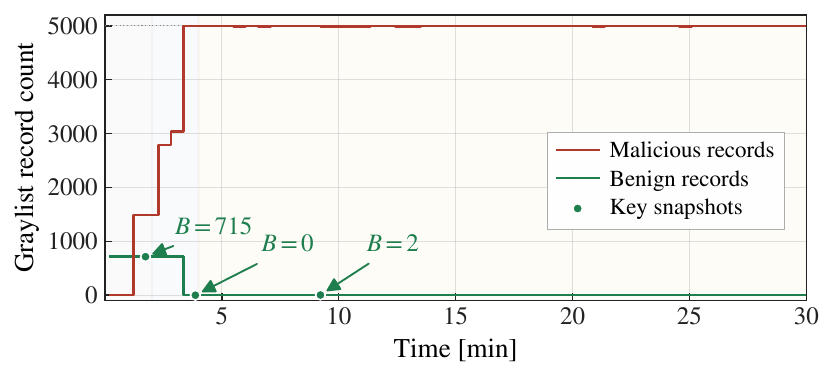}

    \caption{Evaluating \textit{N-II}: graylist infiltration on unreachable target. The 5{,}000-slot graylist over 30\,minutes saturates at minute~4, and only $B=2$ benign records remain by minute~9, below the conservative bound $B_{\max}=144$ (Eq.~\ref{eq:b-max}).}
    \label{fig:nyx_graylist}
\end{figure}

\begin{figure}[t]
    \centering
    \includegraphics[width=\linewidth]{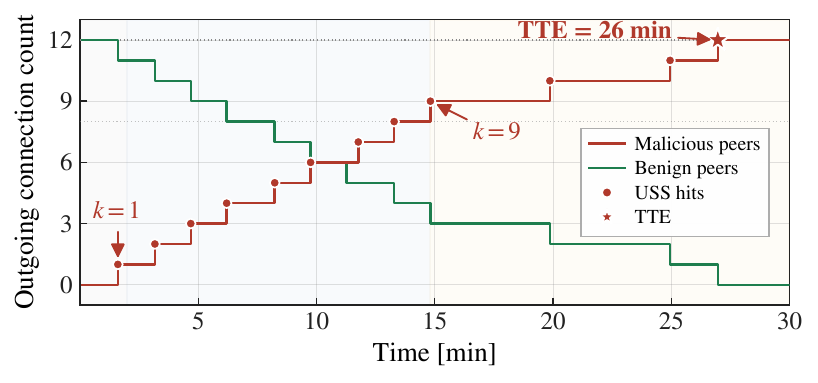}
    \caption{Evaluating \textit{N-III}: outgoing connection takeover. Nyx reaches full takeover in 27 min, replacing all 12 outbounds after 17 \texttt{update\_sync\_search()} rounds with 94.1\% hit rate.}
    \label{fig:nyx_outgoing}
\end{figure}

The goal of \emph{N-II} is to raise the malicious fraction of the
target's graylist to at least $\mathrm{IR}_{\min} \approx 97.1\%$
(Eq.~\ref{eq:ir-min}), providing a high $P_k$ for
subsequent \textsc{uss} replacements.
We sample the target's graylist and whitelist via RPC every
30\,s and log every
\path|gray_peerlist_housekeeping()| invocation.

\heading{Graylist infiltration.}
Before the attack, the target's graylist holds 717 benign entries
(Table~\ref{tab:pre_attack_state}).
Once the attack starts, malicious records are continuously written
into the target's graylist through timed sync.
By minute~2, the graylist has accumulated approximately 1{,}492
malicious records; since the graylist has not yet reached its
5{,}000-entry capacity, new records are appended to the tail
without triggering eviction, and the benign count remains at
approximately~715.
At minute~4, the graylist reaches capacity.
As subsequent malicious records continue to arrive, the
pre-existing benign entries are displaced from the head.
The benign count drops from 715 to~0, and the malicious fraction
reaches 100\% for the first time, exceeding $\mathrm{IR}_{\min}$.
A snapshot at minute~9 shows 4{,}993 malicious and 2~benign records
(malicious fraction 99.96\%), with the benign count well below
$B_{\max} = 144$ (Eq.~\ref{eq:b-max}).
These two benign records correspond to honest peer addresses
reintroduced by the target's remaining benign outgoing connections
during subsequent timed sync rounds.

The experimentally observed $B=2$ falls well below the conservative
theoretical bound $B_{\max}=144$ (Eq.~\ref{eq:b-max}).
We attribute this gap to two factors: (i)~during the early timed-sync
rounds, port-diverse filling returns almost no benign record to the
target, so the target's behavior follows the optimistic case described
in \S\ref{sec-n2}; (ii)~once $k\geq1$, the established malicious
outgoing connections themselves begin streaming fresh malicious
records into the graylist, further evicting the residual benign
entries.

\heading{Whitelist infiltration.}
Within the \emph{N-II} window (first 9\,min),
\path|gray_peerlist_housekeeping()| executes approximately
9~times, selecting malicious records on 8~invocations.
The single benign selection occurs before \emph{N-I} completes its
first filling round, when the graylist malicious fraction is below
70\%.
After \emph{N-I} reaches steady state, all subsequent invocations
select malicious records, continuously injecting attacker-controlled
entries into the whitelist via the Gray$\to$White path.

\subsubsection{\underline{Evaluating \textit{N-III}} (outgoing connection takeover)}
\label{subsec-eval-nyx-phase3}

\emph{N-III} leverages the protocol's autonomous
\path|update_sync_search()| to replace the target's benign outgoing
connections with attacker controlled ones.
We evaluate whether the replacement process converges to $k = 12$
as predicted by the probability model in
\S\ref{subsec-nyx-phase3}, and whether the resulting eclipse state
persists.

\heading{Connection takeover.}
After \emph{N-III} starts, the target reaches CTR${} = 100\%$
(12/12) at approximately minute~27.
During this process, the target undergoes 17
\path|update_sync_search()| (hereafter \textsc{uss}) replacement
rounds, of which 16 select malicious peers (hit rate 94.1\%).
The takeover unfolds in three stages (shaded regions in \figref{fig:nyx_outgoing}).

\textit{Warm-up (0--2\,min).}
This stage corresponds to the window before the first \textsc{uss}
replacement.
\emph{N-I} is still in its first filling round and has not yet
saturated all reachable nodes' whitelists, so the target's graylist
is only partially poisoned.
The first \textsc{uss} replacement fires at approximately
minute~2, when the graylist contains 1{,}492 malicious and
715~benign records (malicious fraction 67.6\%).

\textit{Rapid takeover (2--15\,min).}
Once the target establishes its first malicious outgoing connection,
the takeover enters a positive-feedback stage.
Each attacker-controlled outgoing connection continuously returns
fresh malicious addresses in subsequent timed sync responses,
thereby driving $P_k$ upward.
At minute~4, sustained injection from \emph{N-II} pushes the
target's graylist to its 5{,}000-entry capacity, and $B$ drops to~0
for the first time.
Thereafter, the target's few remaining benign outgoing connections
reintroduce a small number of honest addresses through timed sync,
causing $B$ to fluctuate around 2--3.
These benign records constitute a negligible fraction, so $P_k$
remains close to~1.
Whenever \textsc{uss} releases a benign connection, the replacement
is almost certain to select a malicious peer, driving CTR from
$k = 1$ to $k = 9$.

\textit{Boundary convergence (15--27\,min).}
Once $k \geq 9$, $q_k$ rises to 0.75--0.92.
\textsc{uss} is now more likely to drop an already attacker-controlled
connection than a remaining benign one.
CTR therefore advances only when \textsc{uss} drops a benign
connection and the replacement is again malicious.
Since $P_k$ remains close to~1, $P_k > q_k$ still holds.
The final transition from $k = 11$ to $k = 12$ completes within a
single replacement round, consistent with the boundary analysis in
\S\ref{subsec-nyx-phase3}
($P_{11} \to 1 > q_{11} = 11/12$).

\heading{Whitelist saturation.}
Let $T_0$ denote the start of \emph{N-I}. Both whitelist write paths
produce only malicious entries:
(i)~\textsc{uss} establishes exclusively malicious outgoing
connections, whose \texttt{last\_seen} is continuously refreshed by
timed sync to keep them at the whitelist head;
(ii)~\path|gray_peerlist_housekeeping()| promotes records from the
already malicious-dominated graylist.
By $T_0 + 17$\,min, 12 of the TOP-20 slots are malicious; by
$T_0 + 33$\,min, all 20 are (Appendix, Fig.~\ref{fig:nyx_top20}).
The remaining whitelist slots are filled through
\path|gray_peerlist_housekeeping()|, which selects malicious
graylist records in 1{,}073 of 1{,}074 promotions.
By $T_0 + 18.5$\,h, all 1{,}000 whitelist slots are
attacker-controlled (Appendix, Fig.~\ref{fig:nyx_whitelist_long}).

\heading{Eclipse stability.}
After CTR reaches 100\%, the target enters a 17\,h~47\,min steady-state window in which \path|update_sync_search()| fires 630~times.
Of these, 626 (99.4\%) select malicious replacements and 4 (0.6\%) select benign peers, causing CTR to briefly drop from~12 to~11 before recovering, as the malicious peer pool rapidly displaces the benign candidate.
From minute~85 onward, CTR permanently holds at~12/12 with no further dips.
In the final attack state, all 12~outgoing connections, all 5{,}000~graylist records, and all 1{,}000~whitelist records are controlled by the attacker.
Since the target is unreachable and cannot accept incoming connections, no honest node can independently establish a connection to break the eclipse.
\subsection{Evaluation of the \textsc{Moros} Attack}
\label{subsec-eval-moros}
\begin{figure}[t]
  \centering
  \includegraphics[width=\linewidth]{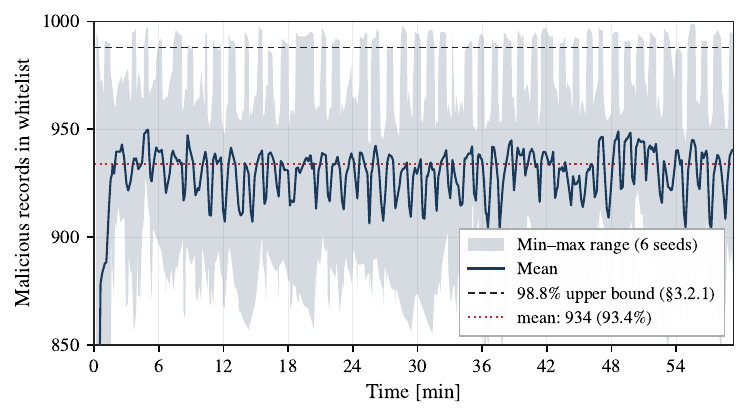}
  \caption{Evaluating \textit{M-I}: seed-node whitelist poisoning. Across the six official seed nodes, malicious records quickly stabilize 934 entries on average, with a 93.4\% occupation rate.}
  \label{fig:seed-whitelist}
\end{figure}

\subsubsection{\underline{Evaluating \textit{M-I}} (seed peerlist poisoning)}
\label{subsec-eval-moros-phase1}

We evaluate whitelist poisoning against six official seed nodes, which serve as the primary bootstrap entry points for newborn nodes.
We deployed 1{,}000 distinct public IPs to initiate handshake requests and monitored the whitelist at 2-second intervals over a one-hour duration.
As illustrated in Fig.\ref{fig:seed-whitelist}, the malicious-record count across the six seed nodes converges within minutes and remains stable around a mean of 934 records (93.4\% OR); the shaded min--max band reflects per-seed variation.

\textit{Factors affecting poisoning rate.}
As observed in Fig.\ref{fig:seed-whitelist}, the whitelist poisoning on seed nodes does not reach the ideal level (i.e., 988 malicious records), and the number of malicious records exhibits noticeable fluctuations.
By inspecting the poisoning logs, we identify two main causes.
First, during periodic timed sync with their existing outgoing peers, seed nodes continuously refresh and retain the corresponding benign peer records in the whitelist.
Second, because seed nodes are hard-coded in the Monero codebase, they frequently receive incoming connection requests from a large number of benign nodes.
Together, these effects induce continuous churn in the whitelist entries, causing the malicious record count to fluctuate around a high level rather than remaining stable.

\textit{Security implications.} A 93.4\% occupation rate is sufficient to undermine the integrity of the bootstrapping process. Statistically, with an OR of 93.4\%, a newborn node fetching 250 records from a seed node would receive an average of $\sim$234 malicious peer records. This high expectation demonstrates that seed nodes can no longer fulfill their role as honest facilitators, exposing newborn nodes to a severe risk of being eclipsed immediately upon initialization.

\subsubsection{\underline{Evaluating \textit{M-II}} (bootstrapping peerlist hijack)}
\label{subsec-eval-moros-phase2}
\begin{figure}[t]
    \centering
    \includegraphics[width=\columnwidth]{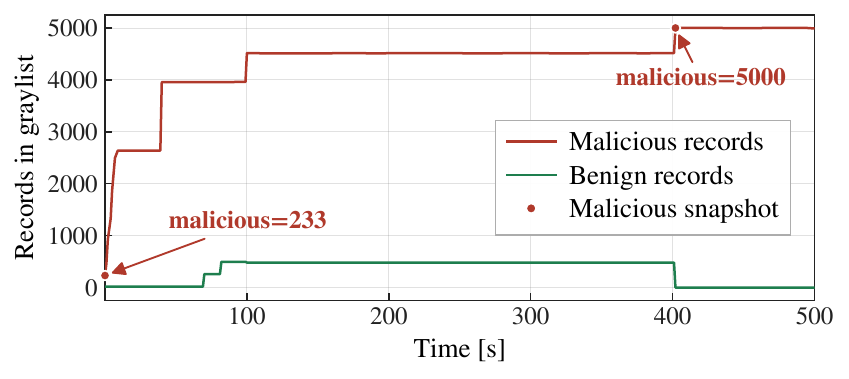}
    \caption{Evaluating \textit{M-II}: bootstrapping-peerlist hijack. Newborn target receives 233/250 malicious seed records by poisoned responses. Its graylist becomes fully malicious in 6min.}
    \label{fig:moros-graylist}
\end{figure}

After continuously launching the whitelist poisoning against the seed nodes for one hour, we start a newborn unreachable node and monitor its outgoing peers and peerlist state via remote RPC with a 1\,s sampling interval.
Fig.~\ref{fig:moros-graylist} shows how the numbers of malicious peers and benign peers in the newborn node's graylist evolve over time.

\heading{Graylist infiltration.}
As shown in Fig.~\ref{fig:moros-graylist}, among the 250 records initially obtained from the seed node, 93.2\% are malicious peers (233/250), while only 6.8\% are benign peers (17/250). With such a high malicious fraction in the graylist from the very beginning, the node is highly likely to select malicious peers for its initial 12 outgoing connections.

Once malicious outgoing connections are established, the attack enters an amplification stage. The target's graylist becomes dominated by malicious peers within only 6 minutes (Fig.~\ref{fig:moros-graylist}). As a result, benign peers have almost no chance to be promoted to the whitelist through the housekeeping mechanism.

During 11 hours of continuous monitoring, the target's graylist remains 100\% malicious OR. This demonstrates that, on the real mainnet, eclipsing newborn unreachable nodes is both immediate and persistent: once the bootstrapping inputs are poisoned, the node is unlikely to recover through the protocol's built-in peer discovery mechanisms.

\begin{figure}[t]
    \centering
    \includegraphics[width=\columnwidth]{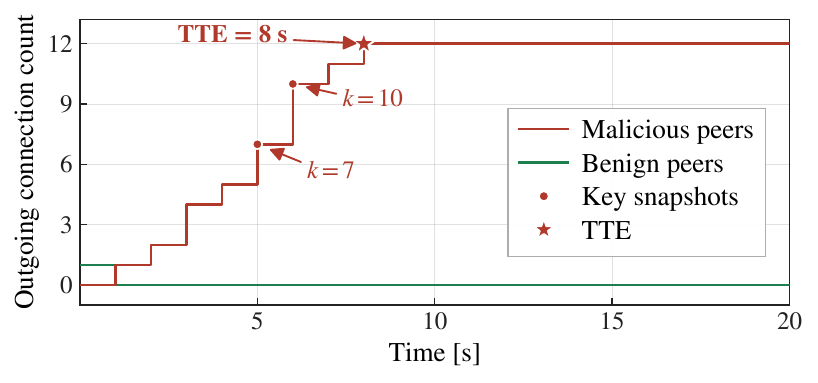}
    \caption{Evaluating \textit{M-III}: outgoing connection takeover on a newborn target. Moros reaches full eclipse in 8 s, after which all 12 outgoing connections are attacker-controlled.}
    \label{fig:moros-outgoing}
\end{figure}

\subsubsection{\underline{Evaluating \textit{M-III}} (outgoing connection takeover)}
\label{subsec-eval-moros-phase3}

\emph{M-III} completes outgoing connection takeover for the newborn: initial selection of all 12~outgoing peers from the poisoned graylist, then reinforcement through Monero's \path|update_sync_search()| cycle (\S\ref{sec-moros}).
We report connection takeover, long-run stability, and peerlist (whitelist) evolution.
For outgoing takeover, we zoom into the first 15\,s after the target node starts (Fig.~\ref{fig:moros-outgoing}).

\heading{Connection takeover.}
During startup, the target first sends handshake messages to a seed node to obtain the initial peerlist.
Since \textit{M-I} has already filled the seed node's whitelist, 233~out of the 250~returned records are malicious peers (93.2\%).
As a result, when selecting outgoing connections, the target has a very high probability of choosing malicious peers from its graylist.
The number of malicious outgoing peers increases in a stepwise manner, passes the annotated states $k=7$ and $k=10$, and reaches 12/12 at $\mathrm{TTE}=8$\,s (Fig.~\ref{fig:moros-outgoing}).
At this point, all outgoing peers are attacker-controlled, and the target enters an eclipse state.

\heading{Eclipse stability.}
\textsc{Moros} maintains strong eclipse stability over 650\,min (${\sim}10.8$\,h) of continuous monitoring after $\mathrm{TTE}$.
Since the initial 12~outgoing peers are attacker-controlled, these malicious peers leverage the protocol's peerlist propagation to keep sending additional malicious records to the target.
Although a benign outgoing connection is observed during this period, it is quickly eliminated as malicious peers continue to dominate the graylist.
These malicious records further push out the remaining benign peers in the peerlist until it is fully dominated.
Consequently, the target has little chance to establish persistent connections to benign peers and remains eclipsed for an extended period.

\heading{Peerlist saturation.}
The newborn node's whitelist becomes dominated by malicious peers within the 1.5\,h attack window (Fig.~\ref{fig:moros-whitelist-app}).
Within 8~seconds after startup, the number of malicious peers in the whitelist quickly rises from~0 to~12.
This is because the target has a very high probability of selecting malicious peers from its graylist, and these malicious peers are easily chosen as outgoing peers. At 63\,s, we observe that the target temporarily establishes one benign outgoing connection, which adds one benign peer to the whitelist.
This observation validates the non-ideal-case analysis: even when a benign connection is temporarily formed, the protocol's maintenance and peerlist dynamics still drive the node toward full malicious takeover.

\section{Discussion}
\label{sec-disscu}

We discuss the following two aspects.

\subsection{Attack Feasibility}
\label{subsec-discussion-feasibility}

This section analyzes how the attacker's IP distribution affects the effectiveness of the eclipse attack. Monero's primary defense is strict /24 subnet filtering in the outgoing peer-selection routine, implemented by \path|make_new_connection_from_peerlist()|.

\heading{Bypassing the subnet filter.}
Unlike earlier studies that focused on Class B ranges, the current Monero protocol (v0.18.4.3) refines filtering to the /24 subnet level. Since an unreachable node maintains 12 outgoing peers, the attacker must control IP addresses spanning at least 12 distinct /24 subnets to avoid being filtered out during candidate selection.
\begin{packeditemize}
    \item  \emph{Resource-sufficient case.}
When the attacker controls 1{,}000 IPs distributed across distinct /24 subnets, malicious peers still dominate the candidate set even after filtering.
    \item  \emph{Resource-limited case.}
If the attacker's IPs are concentrated within only a few subnets, Monero's filtering logic treats many malicious records as duplicates and discards them. In this case, even if the peerlist is fully filled, the proportion of malicious peers in the final candidate set drops significantly.
\end{packeditemize}

\heading{Feasibility of the attack.}
From a cost perspective, acquiring 1{,}000 IP addresses across distinct subnets is affordable. Modern cloud providers and proxy pools enable attackers to assemble a sufficiently diverse node set at low cost. Combined with our lightweight \texttt{py-levin} implementation, a single workstation suffices to mount an eclipse attack at network scale. This shows that Monero’s current subnet-diversity defense is insufficient to prevent targeted attacks against unreachable nodes.

\heading{Coverage robustness.}
The $\mathrm{IR}_{\min}$ bound in \S\ref{sec-n2} (Eq.~\ref{eq:ir-min}) assumes that \textit{N-I} directly poisons every reachable node in the network. In practice, the attacker's reachable-node inventory may incompletely cover the network due to node churn or probing latency. Let $c_0$ denote the fraction of reachable nodes directly poisoned by \textit{N-I}.

A reachable node not directly poisoned by \textit{N-I} is not safe: it still receives timed sync responses from its own outgoing peers, and once those peers' whitelists are dominated by attacker records, the indirectly targeted node ingests poisoned records into its graylist. Through the Gray$\to$White promotion path (\S\ref{sec-background-p2p}), these records progressively enter its whitelist as well. In effect, an unpoisoned reachable node behaves as an unreachable target with respect to peerlist contamination, with the same protocol-driven infiltration mechanism described in \S\ref{sec-n2} propagating attacker records into it.

The effective poisoning coverage $c_{\mathrm{eff}}$ converges toward 1 over time even when $c_0 < 1$. The $\mathrm{IR}_{\min}$ graylist pollution lower bound (Eq.~\ref{eq:ir-min}) holds asymptotically irrespective of \textit{N-I}'s direct coverage, provided $c_0$ is large enough to seed the propagation. We do not empirically measure $c_0$ or the convergence rate of $c_{\mathrm{eff}}$. Characterizing this propagation dynamic is beyond our current commitment.

\subsection{Generalizability}
\label{subsec-generalizability}

\heading{From emulation to mainnet.}
Although our \textsc{Nyx} evaluation uses a 1{,}200-node Monero network, the local convergence mechanisms of \textit{N-II} and \textit{N-III} do not depend on the global network size.
Both phases act on the target's local peerlist rather than the global peerlist.
The Monero client maintains a 1{,}000-entry whitelist and a 5{,}000-entry graylist; neither capacity grows with the number of reachable mainnet nodes.
Therefore, as long as \textit{N-I} continuously poisons the reachable nodes connected to the target's current outgoing peers, the target's graylist still converges under \textit{N-II} to a state whose malicious fraction is at least $\mathrm{IR}_{\min}$.
In that state, \textit{N-III} connection-takeover analysis still applies.

The local convergence above is driven by a positive-feedback loop inside the target.
Once a malicious peer is selected as an outgoing connection, it becomes a persistent source of fresh malicious records during timed sync.
This reduces the remaining benign candidate set in the graylist and increases the probability that later replacement rounds also select malicious peers.
The same asymmetry applies to the whitelist: unreachable nodes cannot receive benign inbound insertions, while attacker-controlled outgoing peers keep refreshing their \texttt{last\_seen} timestamps and malicious graylist entries continue to be promoted through housekeeping.

\textbf{The main boundary for mainnet deployment lies in \textit{N-I} coverage maintenance.}
\textsc{Nyx} requires the attacker to maintain a reachable-node inventory and repeatedly fill that set so that the target's current outgoing peers are covered. Node churn and failures increase this cost and may delay convergence, but they do not change the local \textit{N-II}/\textit{N-III} conditions once sufficient poisoning is reached. For ethical reasons, we do not run \textsc{Nyx} against arbitrary existing unreachable mainnet nodes; our mainnet claim is mechanism transferability, not full mainnet validation.

\heading{Beyond Monero.} Our attack exposes a general failure mode affecting \emph{outbound-only} P2P nodes that rely on transitive peer propagation and locally bounded peerlists. Any system in which (i) nodes cannot accept inbound connections, (ii) peer discovery is learned exclusively from existing neighbors, and (iii) connection refresh is driven by local peer state rather than global diversity guarantees, may be vulnerable to similar eclipse dynamics. While our evaluation focuses on Monero, these conditions arise naturally in many NATed or firewalled deployments. The attack surface demonstrated is not specific to a single cryptocurrency network.

\section{Countermeasures}
\label{sec-countermeasures}

This section discusses why natural defenses are insufficient  (\S\ref{subsec:counter-why-fail}) and what changes are needed to mitigate the attacks (\S\ref{subsec:counter-protocol}). We also discuss the remaining structural challenge (\S\ref{subsec:counter-structural}).

\subsection{Why Existing Defenses Fail}
\label{subsec:counter-why-fail}

Monero's deployed defenses cannot block the attack under our threat model.
The anchor list requires a node to reuse recent outgoing peers after restart, yet Nyx does not rely on restart and a Moros-targeted newborn node has an empty anchor list.
The \texttt{v0.18.3.2} double-spend connection retention patch addresses active connection reset, yet Nyx and Moros exploit the autonomous replacement in \path|update_sync_search()|.
The \texttt{v0.18.4.3} /24 subnet diversity filter requires outgoing connections to span distinct /24 subnets, yet our threat model assumes attacker coverage of more than $1{,}000$ /24 subnets, so the filter only raises cost by a constant factor.
The Erebus-inspired proposal to increase outgoing connections is equally ineffective, since the OR upper bound in \S\ref{sec-n1} varies with $|A| - N_{\mathrm{out}}$ and the attacker maintains saturation by expanding its IP budget accordingly.
These defenses share a common limitation in that each targets a specific attack technique without addressing the Sybil-resource precondition.
We make this constant-cost property explicit through a six-group instantiation that caps the adversarial budget at $6{,}000$ node-identities independent of network size, yet still achieves a complete eclipse; see Appendix~\ref{app:sixgroup}.
\subsection{Protocol-Level Modifications}
\label{subsec:counter-protocol}

Our whitelist-filling method follows prior work on Monero~\cite{shi2025eclipse}.
Previously discussed countermeasures include preventing whitelist insertion through incoming handshakes and increasing whitelist capacity.
Following our disclosure, Monero developers proposed a fix that restricts peerlist exchange and graylist insertion.
We discuss how these changes limit peerlist enumeration, graylist filling, and multi-port amplification, along with their costs to legitimate nodes.

\heading{Limiting peerlist responses.}
In \textit{N-I-1}, \textsc{Nyx} establishes connections and repeatedly sends timed sync requests to enumerate reachable nodes' whitelists.
The attacker checks which malicious records have been inserted and repeats handshakes from missing attacker IPs to complete the filling.

The fix limits responses in two ways.
First, each connection can return at most 250 records in a 10-minute window, counting handshake and timed sync responses together.
This bounds the number of whitelist records an attacker can obtain through repeated requests on one connection.
Second, the node caches a snapshot of up to 1{,}000 whitelist records.
Peerlist responses select records in snapshot order, skipping those already sent over the connection.
While the snapshot remains unchanged, new connections receive the same initial batch in their handshake responses.
Repeated connections therefore do not yield fresh whitelist samples.
Once the snapshot is at least 10 minutes old, the node refreshes it when a peerlist response is next needed.
Snapshot updates are independent of each connection's counting window.

Together, these limits make whitelist enumeration and filling verification in \textit{N-I-1} less efficient.
The response limit also slows the forwarding of poisoned records from honest reachable nodes to the target in \textit{N-II}.
The tradeoff is that legitimate nodes may need more time to obtain additional candidates.
New whitelist records must also wait for a snapshot update before they can be returned.

\heading{Limiting peerlist reception.}
In \textit{N-I-2}, attacker nodes maintain connections with reachable nodes and repeatedly return unreachable trash records in timed sync responses.
These records evict benign graylist entries, reducing their chances of being probed and promoted to the whitelist.
This helps sustain whitelist poisoning.

The node counts received records separately for each connection and processes at most 250 records per 5-minute window.
Excess records are discarded before graylist insertion.
The node also rejects timed sync requests and closes the connection if the handshake has not completed, as indicated by the absence of a valid \texttt{peer\_id}.

The reception limit reduces the number of trash records that an attacker can insert through successive timed sync responses on one connection, slowing graylist filling in \textit{N-I-2}.
It also applies to peerlists received by an unreachable target from its outgoing peers, limiting continued graylist injection in \textit{N-II}.
Legitimate neighbors face the same restriction, which may slow normal peer discovery.

\heading{Limiting graylist records from the same IP and subnet.}
\textsc{Nyx} assigns different ports to the same attacker IP across reachable nodes' whitelists.
In \textit{N-II}, the resulting $\langle\mathrm{IP},\mathrm{port}\rangle$ records propagate through peerlist exchanges into the unreachable target's graylist as distinct entries.
This amplifies the attacker's record count.

Before inserting a new graylist record, the node checks its IP address.
If the graylist already contains that IP, records using other ports are ignored.
This check preserves the underlying $\langle\mathrm{IP},\mathrm{port}\rangle$ index.
The node also rejects new records from an IPv4 /16 subnet once the graylist contains 64 records from that subnet.

The same-IP check removes the multi-port amplification used in \textit{N-II}, while the /16 check limits records from each subnet.
To insert the same number of malicious records into the target's graylist, the attacker must control more distinct IPs spread across more /16 subnets.
This increases the required address resources.
The tradeoff is that only one record from legitimate Monero instances sharing a public IP can pass the insertion check.
Further legitimate records from a /16 subnet are also ignored once it reaches 64 graylist entries.

The reception and graylist insertion limits also constrain \textsc{Moros}'s ability to sustain peerlist poisoning.

\subsection{A Structural Perspective}
\label{subsec:counter-structural}

These measures restrict specific peerlist poisoning paths.
They do not, by themselves, guarantee protection against a persistent Sybil adversary under our threat model.
For unreachable nodes, restricting record reception, propagation, and insertion does not provide honest peer information independent of the current peerlist.

\heading{Why local monitoring is insufficient.}
Monitoring local peerlists and outgoing connections can reveal abrupt changes.
However, \textsc{Nyx} gradually takes over connections through normal timed sync and outgoing-connection replacement.
The changes observed by the target still follow the protocol.
A newborn node targeted by \textsc{Moros} has no historical view for comparison: the poisoned seed response defines its initial candidate set.
Local consistency alone therefore cannot establish whether the node remains connected to the honest network.

Long-term defenses must consider how unreachable nodes can obtain peer information independently of their current outgoing peers.
Two directions merit further study.
The first binds peer identities to scarce resources, increasing the cost of maintaining many Sybil identities.
The second uses independent channels to check a node's network view and help detect eclipse states.
Both involve broader protocol design choices beyond the peerlist changes discussed here.

\section{Conclusion}

We present the first practical eclipse attack targeting \emph{unreachable} Monero nodes, a class of nodes previously believed to be resilient due to their inability to accept incoming connections. We show that by combining indirect peerlist poisoning with built-in connection update mechanism, an attacker can reliably take over all outgoing connections and keep victim nodes isolated for extended periods.

We instantiate the attack in two settings: \textsc{Nyx}, which targets \emph{existing} unreachable nodes, and \textsc{Moros}, which targets \emph{newborn} unreachable nodes. In our 1{,}200-node emulation, \textsc{Nyx} reaches full connection takeover (CTR $=$ 100\%, 12/12) within ${\sim}27$\,min after starting, and the eclipse persists throughout an 18.2\,h monitored window. We further validate \textsc{Moros} on the Monero mainnet, demonstrating an \emph{eclipse-at-birth} against newborn unreachable nodes. Ethically, all mainnet experiments were conducted on controlled targets and carefully designed to avoid persistent impact on the network.

\bibliographystyle{ACM-Reference-Format}
\bibliography{bib}

\appendix

\section*{Open Science}
\label{sec:open-science}

We share the full set of code, configurations, and
one complete raw trace required to construct the emulation
environment and reproduce both attacks. Our artifacts are
hosted at:
\textcolor{teal}{\url{https://doi.org/10.5281/zenodo.22683130}}.

We deliberately withhold two pieces of data.
(i)~The raw RPC traces from our Monero \emph{mainnet}
\textsc{Moros} experiment are not released, because they contain
$\langle$IP, port$\rangle$ records of third-party nodes that have
not consented to disclosure. (ii)~The specific 1{,}000 public IPs
used in the mainnet experiment are withheld during the
responsible-disclosure window, to prevent direct reuse of the same
infrastructure against the currently deployed Monero seed nodes.
The released attack code accepts any equivalent IP list as input,
so neither omission affects the reproducibility of the methodology.

\heading{Availability.}
We release the following materials online: (1)~the SEED-Emulator scripts used to generate the 1{,}200-node Monero emulation network, including multi-IP provisioning, the network-impairment injector, and NAT
simulation rules; (2)~the patched Monero v0.18.4.3 client together
with its build files; (3)~the \textit{py-levin}-based whitelist-filling
code and the graylist-filling code; (4)~the RPC-based monitoring
scripts and configuration files; (5)~one complete set of raw
experimental traces from a \textsc{Nyx} emulation run; and
(6)~a top-level README documenting the reproduction workflow.

\section*{Ethical Considerations}
\label{sec:ethical-considerations}

Our whitelist-filling attack against seed nodes is designed to avoid persistently taking over connections of any uncontrolled nodes on the mainnet. The reasons are as follows.

\begin{packeditemize}
    \item \textbf{Our controlled nodes do not provide usable peerlist responses to uncontrolled nodes.}
    Our whitelist-filling program only serves normal protocol interactions (handshake responses and timed sync responses) to \emph{our controlled nodes}. For all other nodes, it only replies to \texttt{PING} messages and quickly closes the connection. As a result, an uncontrolled newborn node cannot obtain a valid handshake response from our controlled peers and thus cannot establish normal outgoing connections to them.

    In the worst case, a seed node may return a peerlist field that contains only malicious peers. The uncontrolled node will then attempt outgoing connections and fail repeatedly. Under Monero's connection management logic, after more than three consecutive failures when trying to connect to candidates from its peerlist, the node falls back to seed nodes again and sends a handshake request to fetch a fresh peerlist field. In our mainnet experiments, we did not persistently occupy the whitelists of \emph{all} seed nodes. Therefore, in subsequent retries, an uncontrolled node can still obtain a non-trivial fraction of benign peers from seed nodes and proceed with normal connection establishment.

    \item \textbf{/24 subnet filtering limits the impact even if malicious peers are returned.}
    Although we control 1{,}000 public IP addresses, they are distributed across only a limited number of /24 subnets (e.g., 8). Since Monero applies /24 subnet filtering during outgoing peer selection, an uncontrolled newborn node cannot establish 12 outgoing connections to our controlled peers, even if the seed node's returned peerlist field is heavily polluted. Therefore, our seed-node whitelist filling does not disrupt the normal operation of other uncontrolled nodes on the mainnet.
\end{packeditemize}

Based on these constraints, our mainnet experiments do not perform eclipse attacks against other uncontrolled nodes. At most, the potential impact is limited to a small number of failed connection attempts to unusable peer records during startup and the corresponding retry overhead.

\section{Deferred Related Work}
\label{sec-relatedwork}


\heading{Eclipse attacks on Bitcoin.}
Heilman et al.~\cite{heilman2015eclipse} first systematized an eclipse attack against Bitcoin nodes: the attacker persistently poisons the target's address database and occupies its incoming connection slots, and then waits for a restart so that the victim forms outgoing connections from the polluted candidate set, eventually becoming isolated. Tran et al.~\cite{tran2020stealthier} later proposed a more stealthy eclipse attack, \emph{Erebus}, which leverages an AS-level adversary. By using AS-owned address resources to infiltrate the target's candidate pool and intercepting/replacing TCP connections between the victim and its neighbors at the network layer, the attacker naturally becomes a man-in-the-middle and can progressively gain full control. Tran et al.~\cite{tran2021routing} further proposed an integrated defense framework to mitigate the impact of the Erebus attack. In addition, the Bitcoin Core community has noted that eviction-style connection management can be susceptible to eclipse attacks~\cite{website:bitcoin}.

\heading{Eclipse attacks on Ethereum.}
Marcus et al.~\cite{marcus2018low} exploited a design weakness in early versions of the geth client: the client starts its UDP listener (accepting unsolicited \texttt{PING} messages and populating its table) before entering the seeding process. An attacker can use Sybil nodes to pre-fill the victim's discovery table, thereby influencing the victim's outgoing peer choices. Henningsen et al.~\cite{henningsen2019eclipsing} further took advantage of geth v1.8.0's outgoing-selection behavior, where the client chooses the head entry from each bucket (the discovery table contains 17 buckets). By sending unsolicited \texttt{PING} messages to place attacker-controlled nodes at bucket heads, the adversary can steer outgoing connections. Starting from geth v1.9.0, outgoing peers are chosen uniformly at random from the peerlist, which renders these head-entry-based eclipse techniques ineffective. Shi et al.~\cite{shi2026eclipse} presented an end-to-end implementation of an eclipse attack against Ethereum execution-layer nodes, showing that an adversary can exploit Ethereum’s bootstrapping and peer-management logic to fully isolate a victim upon restart.

\heading{Eclipse attacks on Monero.}
A common limitation of many prior eclipse techniques is that the attacker cannot precisely control \emph{when} the eclipse takes place and often has to wait for a victim restart. Moreover, such techniques may fail against networks with proactive countermeasures. In Monero, for example, the lack of an eviction mechanism \cite{shi2025eclipse} prevents an attacker from actively expelling honest connections, and the absence of a strict cap on incoming connections makes it difficult to block new honest incoming peers. In contrast, Bitcoin's eviction policy \cite{heilman2015eclipse,website:bitcoin} and Ethereum's incoming-connection \cite{shi2026eclipse} limits can be more readily ``leveraged'' by attackers to complete an eclipse. Shi et al.~\cite{shi2025eclipse} proposed the \emph{controllable} eclipse attack, enabling an attacker to launch an eclipse at an arbitrary time and complete it within minutes. The work introduced \emph{connection-reset attacks} that exploit a node's internal mechanisms to trigger connection updates and conducted an eclipse attack on the Monero network.

\heading{Partition attacks.}
Partition attacks split a blockchain network into two or more mutually isolated components~\cite{tran2020stealthier,ha2023sustainability,chen2025dissecting}, preventing normal communication across partitions. From the perspective of outcomes, partitioning can be viewed as a generalization of eclipse attacks, i.e., a special case of partitioning that targets a single victim (or a small set of victims). Muhammad et al.~\cite{saad2021syncattack} proposed \emph{SyncAttack} against Bitcoin: by occupying a node's incoming connections and controlling the incoming and outgoing paths to that node, the attacker can create an adversary-controlled partition that prevents (even mining) nodes from connecting to honest peers, potentially enabling transaction deanonymization. Heo et al.~\cite{heo2023partitioning} proposed \emph{GethLighting} against Ethereum: by controlling roughly half of the victim's TCP connections and launching low-rate DoS traffic, an attacker can disrupt block synchronization, stall the victim's block height, and separate it from the Ethereum network. Muhammad et al.~\cite{saad2023three} further proposed an efficient partitioning technique that can simultaneously affect multiple interdependent cryptocurrency networks by carefully crafting network traffic and node behaviors to split targets into multiple non-communicating sub-networks.

\heading{Unreachable nodes attacks.}
Wang et al.~\cite{wang2017towards} conducted one of systematic studies of unreachable peers in the Bitcoin network. By deploying a large number of monitoring nodes, they showed that unreachable nodes exhibit distinct connection patterns and transaction propagation behaviors, which make them vulnerable to deanonymization attacks such as the first-spy estimator~\cite{bojja2017dandelion}. Biryukov et al.~\cite{biryukov2014deanonymisation} demonstrated that unreachable nodes can be uniquely fingerprinted through the set of peers they advertise via ADDR messages, allowing multiple transactions created within the same session to be linked to the same node. Biryukov et al.~\cite{biryukov2015bitcoin} further showed that unreachable nodes connecting through Tor can be deanonymized across multiple sessions. Mastan et al.~\cite{mastan2017new} exploited block request patterns to identify unreachable nodes over time, enabling session linking that can be combined with other deanonymization techniques. 

Heimbach et al.~\cite{heimbach2025deanonymizing} studied privacy leakage in Ethereum’s P2P network. By observing the propagation of consensus attestations and exploiting the fact that a peer’s message pattern reveals which validators it hosts, they can link validator identities to IP addresses, deanonymizing a non-trivial fraction of validators. Shi et al.~\cite{shi2026proxymark} proposed \emph{ProxyMark} to deanonymize Monero transactions over Tor. For Tor clients, it combines indirect peerlist poisoning, outbound connection occupation, and traffic watermarking to link transactions to source IP addresses. Our work focuses on sustained isolation of unreachable Monero nodes by taking over all 12 public-network outgoing connections.

Network-layer attacks in cryptocurrency P2P networks~\cite{chen2020survey,neudecker2018network} span topology inference, deanonymization, and routing-level manipulation. 
Topology inference either passively leverages propagation artifacts and timing signals (e.g., address dissemination and transaction arrival-time variations), which can be noisy and brittle under protocol/client changes, or actively probes adjacency by injecting crafted transactions (e.g., orphan-based probes or double-spend techniques) at non-trivial cost; simulation-based mapping is also used but may deviate from the real topology \cite{miller2015discovering, cao2020exploring, delgado2019txprobe, grundmann2018exploiting}. 
Network-layer deanonymization aims to link participants (e.g., validators or transaction originators) to network identifiers under realistic observation capabilities \cite{klusman2018deanonymisation,lin2024denseflow, wang2024deanonymizing,zhou2022behavior,heilman2015eclipse,heimbach2025deanonymizing}. 
AS-level adversaries can exploit BGP hijacking or passive routing visibility to manipulate or monitor P2P connectivity (e.g., partitioning, delaying propagation, and enabling network-layer deanonymization), providing capabilities largely orthogonal to protocol-specific exploits \cite{zhang2026ntssl, doumanidis2026routing,apostolaki2017hijacking, apostolaki2019sabre, apostolaki2021perimeter, sun2015raptor, birgelee2018bamboozling}.

\section{Technical Background}
\label{apdx-background-p2p}

We provide the protocol details underlying the attack. Every mechanism described is exploited in at least one attack phase (\S\ref{sec-attack}).

\subsection{P2P Communication}
\label{app:p2p}

Monero nodes communicate via three message types.

\heading{Handshake messages.} When a node initiates an outgoing connection, it establishes a TCP connection to the target \texttt{<IP:Port>} and sends a handshake request. The remote peer responds with a handshake response containing a \emph{peerlist} field of up to 250~peer records randomly sampled from its whitelist. Upon receiving this response, the connection is established. For incoming connections, the receiving node asynchronously sends a \texttt{PING} to the connecting peer's claimed \texttt{<IP:Port>}. Only if it receives a valid \texttt{PONG} does it treat the peer as reachable and insert it into its whitelist; otherwise it does not whitelist the peer. An attacker can exploit the handshake mechanism to inject peer records into a target's whitelist by initiating incoming connections that pass \texttt{PING}/\texttt{PONG} validation.

\heading{Timed sync messages.} Every ${\sim}60$~seconds, a node sends timed sync requests to all its neighbors. Each response carries a peerlist of up to 250~peer records sampled from the responder's whitelist. The \texttt{last\_seen} timestamps in these records are cleared before sending to prevent topology inference. Because a node sends timed sync requests to all neighbors every minute, the attacker receives repeated opportunities to inject peer records into the target's graylist.

\heading{Sent-address list.} The sender maintains a per-connection \emph{sent-address list} that records which \texttt{<IP:Port>} entries have already been included in previous responses. When constructing a peerlist for a timed sync response, the sender randomly selects up to 250~peer records from its whitelist and removes those already present in the sent-address list. Only the remaining peer records are included in the response. Thus, the set of new peer records a single connection can deliver shrinks with each timed sync round and is ultimately bounded by the sender's whitelist size. Once the entire whitelist has been transmitted, subsequent responses carry no new peer records. This property bounds how many benign peer records any single honest outgoing connection can contribute to the target's graylist.

\subsection{Peerlist Management}
\label{app:peerlist}

Each node maintains a local peer database organized into three lists: the \emph{anchorlist}, \emph{whitelist}, and \emph{graylist}. We describe each list and the properties that our attack exploits.

\subsubsection{\underline{Whitelist}}
\label{app:whitelist}

The whitelist stores peers whose reachability has been verified. It holds up to 1{,}000 entries, sorted by \texttt{last\_seen} (most recent communication time) in descending order.

\heading{Insertion.} A node inserts a peer's \texttt{<IP:Port>} into its whitelist through three paths:

(i)~\emph{Incoming-connection validation.} When a node receives a handshake for an incoming connection, it sends \texttt{PING} to the peer's claimed \texttt{<IP:Port>}. If it receives a valid \texttt{PONG}, it inserts the peer into the whitelist with \texttt{last\_seen} set to the current time; if not, it does not add the peer to the whitelist. Whitelist records are indexed by their full \texttt{<IP:Port>} addresses. An attacker can exploit this path by initiating handshakes with many distinct IPs to fill the whitelist rapidly.

(ii)~\emph{Periodic probing promotion (Gray$\to$White).} Every 60~seconds, the node executes \texttt{gray\_peerlist\_housekeeping()}, which randomly samples one peer record from the graylist and attempts a handshake. If the handshake succeeds, the node removes that record from the graylist and inserts it into the whitelist with an updated \texttt{last\_seen}. If the handshake fails, the node deletes that record from the graylist (it is not retained).

(iii)~\emph{Promotion by outgoing connection.} When the node selects a peer from the graylist for an outgoing connection, the peer is promoted to the whitelist and its \texttt{last\_seen} is updated.

\heading{Ordering and \texttt{last\_seen} updates.} Entries are sorted by \texttt{last\_seen} in descending order. The timed sync mechanism updates \texttt{last\_seen} every 60~seconds for outgoing peers that successfully respond. Incoming connections do \emph{not} update \texttt{last\_seen}; this design prevents an attacker from refreshing its ranking through repeated incoming connections alone. As a consequence, a node's current outgoing connections consistently occupy the top positions in the whitelist.

\heading{Eviction.} When the whitelist exceeds its capacity (1{,}000), the entry with the smallest \texttt{last\_seen} is evicted. As current outgoing peers continuously refresh \texttt{last\_seen}, they resist eviction. An attacker must therefore inject enough records to push out all non-outgoing entries before the outgoing peers themselves can be displaced.

\subsubsection{\underline{Graylist}}
\label{app:graylist}

The graylist stores candidate peers whose reachability has not been verified. It holds up to 5{,}000 entries and follows FIFO eviction.

\heading{Insertion.} When a node receives a peerlist (from a handshake or timed sync response), it checks each peer record against its existing whitelist and graylist. Duplicate peer records are silently discarded. New peer records are appended to the tail of the graylist without any reachability check. The \texttt{last\_seen} field of inserted peer records is set to~0 to prevent attackers from manipulating ordering via forged timestamps. Because insertion requires no validation, an attacker who controls an outgoing connection can inject arbitrary peer records into the target's graylist at a rate of up to 250~peer records per timed sync round.

\heading{FIFO eviction.} When new insertions cause the graylist to exceed its capacity, the oldest entries at the head are evicted first. An attacker who continuously injects distinct records forces older entries out of the graylist, reshaping the candidate pool for subsequent connection attempts.

\subsubsection{\underline{Anchorlist}}
\label{app:anchorlist}

The anchorlist records the node's current outgoing connections. Each entry contains the peer's \texttt{<IP:Port>} and a \texttt{first\_seen} field set to the time the outgoing connection was established. When a connection is dropped, the corresponding entry is deleted.
The anchorlist is used primarily during cold start: the node reconnects to anchor peers in ascending order of \texttt{first\_seen}, then clears the remaining entries once a connection succeeds. During continuous operation, the anchorlist reflects the current set of outgoing connections and is rebuilt dynamically as connections are established and dropped.

\subsubsection{\underline{Outgoing Peer Selection}}
\label{app:selection}

A Monero node maintains up to 12~outgoing connections by default. The selection logic is driven by \texttt{connections\_maker()} and depends on the operating stage and the current outgoing-connection count.

\heading{Cold start.} If the whitelist is empty (newborn node), the node connects to a hardcoded seed node, obtains an initial peerlist of up to 250~peer records, and writes them into the graylist. If the whitelist is non-empty, the node first attempts to restore connections from the anchorlist in ascending order of \texttt{first\_seen}. Once an anchor connection succeeds, the remaining anchor entries are cleared, and subsequent connections are filled from the whitelist. A newborn node's initial candidate pool consists entirely of peer records provided by seed nodes. If those peer records are attacker-controlled, the node establishes malicious outgoing connections from the start.

\heading{Steady-state selection.} During normal operation, the node selects candidates from the whitelist or graylist based on the current outgoing-connection count $N_\text{out}$ and a threshold, i.e., \texttt{expected\_wh-} \texttt{ite\_connections} (default:~8, i.e., 70\% of~12):

\begin{packeditemize}
\item \emph{Whitelist-first.} If $N_\text{out} < 8$, the node prioritizes candidates from the whitelist.
\item \emph{Graylist-first.} If $N_\text{out} \geq 8$, the node selects from the graylist. Only after three consecutive graylist failures does the node fall back to the whitelist. This is the dominant path during steady operation.
\end{packeditemize}

Because \path|update_sync_search()| drops exactly one connection (reducing $N_\text{out}$ from~12 to~11, which is still $\geq 8$), the replacement consistently follows the graylist-first path. An attacker who dominates the graylist therefore controls the replacement outcome.

\heading{Deduplication and /24 filtering.} Before selecting a candidate, the node applies two filters to the chosen list:

(i)~\emph{Host-level deduplication.} For records sharing the same IP but different ports, only one representative is retained.

(ii)~\emph{/24 subnet filtering.} The node shuffles the candidates, retains only the first record per /24~subnet, and excludes subnets already represented among existing outgoing peers.

After filtering, graylist selection is approximately uniform random over the remaining candidates. Whitelist selection still biases toward records with more recent \texttt{last\_seen}. These filters require the attacker to distribute malicious peers across many distinct /24~subnets to maintain a high post-filter candidate count.

\subsection{Connection Update Mechanism}
\label{app:update}

The protocol periodically refreshes outgoing connections via \path|update_sync_search()|. This mechanism creates a recurring reconnection entry point that the attacker exploits in \textit{N-III} and \textit{M-III}.

\heading{Trigger condition.} The mechanism fires when three conditions hold simultaneously: (i)~the node is synchronized (local chain height matches the network tip); (ii)~the number of outgoing connections equals the configured maximum (default:~12); (iii)~fewer than two outgoing connections are in \texttt{state\_synchronizing}. For a synchronized node, nearly all outgoing peers remain in \texttt{state\_normal}. The count of \texttt{state\_synchronizing} connections is typically zero and rises briefly when a new block arrives. The trigger therefore fires frequently during normal operation.

\heading{Which connection is dropped.} The node iterates over all managed connections and disconnects the last \texttt{state\_normal}, non-anchor outgoing peer encountered. Because connections are stored in an \texttt{unordered\_map}, the iteration order is effectively random. In practice, the node drops a uniformly random eligible outgoing connection. Even well-behaved benign peers may be removed.

\heading{Replacement.} After dropping a connection, the node immediately triggers its connection-selection logic (\S\ref{app:selection}) to fill the released slot. Because $N_\text{out}$ drops from~12 to~11 ($\geq 8$), the replacement follows the graylist-first path. Each update cycle therefore exposes the peer-selection logic to the attacker's polluted candidate set. Repeating this process over multiple cycles allows the attacker to replace benign outgoing peers one by one.

\subsection{Seed Nodes and Newborn Nodes}
\label{app:seed}

Seed nodes are a small set of reachable-node addresses hardcoded in the Monero client. They serve as initial peer-information sources during node startup.

A node connects to a seed node in two cases: (i)~when the node starts for the first time or its whitelist is empty; (ii)~when repeated attempts to connect to candidates from the whitelist and graylist fail. In both cases, the node sends a handshake request and receives a peerlist of up to 250~peer records, which it inserts into its graylist.

A newborn node begins with empty whitelist, graylist, and anchorlist. Its entire initial candidate pool consists of the records returned by its first seed-node handshake. If an attacker has poisoned the seed node's whitelist, the newborn node's graylist is dominated by malicious records from the start, and all subsequent outgoing connections are established from this polluted pool.

\section{Moros Whitelist Saturation}
\label{app:moros-whitelist-saturation}

\begin{figure}[H]
    \centering
    \includegraphics[width=\columnwidth]{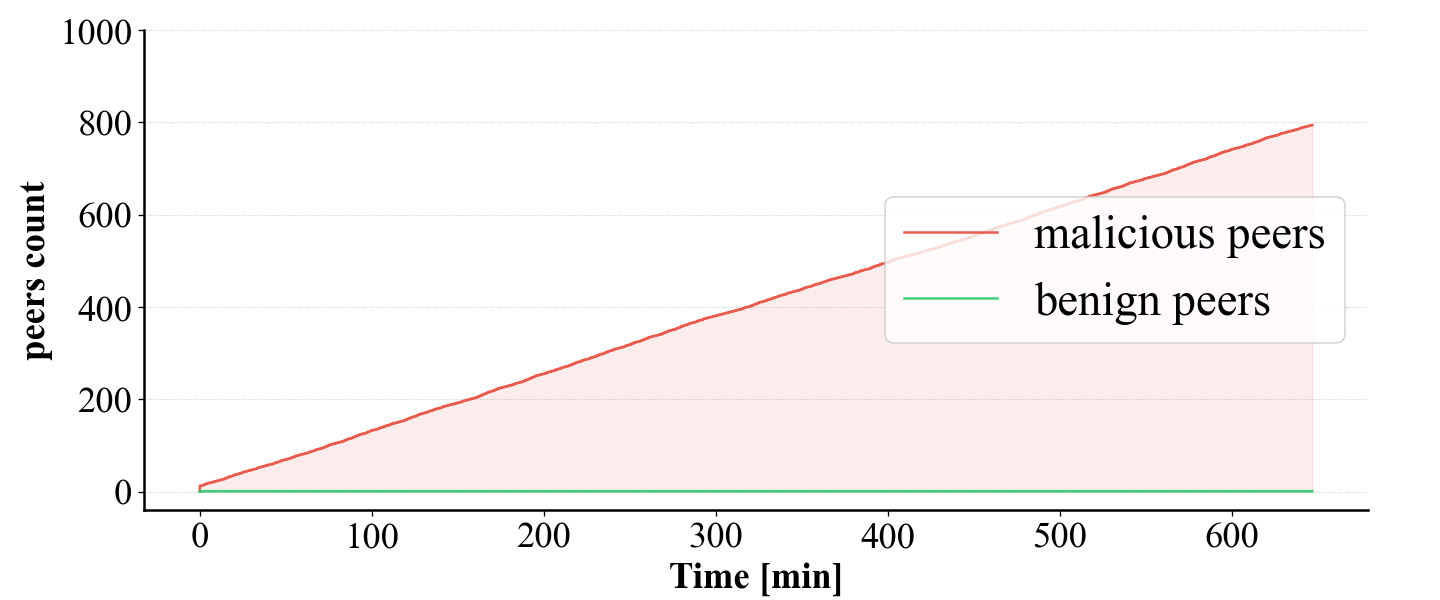}
    \caption{\textsc{Moros} \textit{M-III}: whitelist saturation on the newborn
    target. Malicious peers rapidly enter whitelist after startup, while a
    temporary benign whitelist entry appears at 63\,s.}
    \label{fig:moros-whitelist-app}
\end{figure}

This appendix extends \S\ref{subsec-eval-moros} (peerlist
saturation paragraph) by showing how the newborn target's whitelist
evolves throughout the entire \textsc{Moros} attack.
Fig.~\ref{fig:moros-whitelist-app} shows the malicious and benign
peer counts in the target's 1{,}000-slot whitelist over the
650\,min (${\sim}10.8$\,h) monitoring window.
Within 8\,s of startup, the 12 slots that correspond to the initial
outgoing connections are filled with malicious entries.
At 63\,s, the target briefly establishes a benign outgoing
connection, and a transient benign whitelist entry appears
accordingly.
This benign connection is quickly torn down, and the corresponding
entry is gradually evicted by malicious peers.
For the rest of the run, the malicious count grows monotonically
while the benign count stays at zero.

\section{Detailed Convergence Analysis of \textit{N-III}}
\label{app:niii-convergence}

\textbf{Round-level eviction.}
When a node is selected from the graylist as a new outgoing connection, it is removed from the graylist and added to the whitelist.
Before the next replacement round, this new outgoing peer returns about 500 records to the target through one handshake response and at least one timed sync response.
These records enter the graylist tail and FIFO-evict the same number of records from the head.
If benign records are uniformly distributed in the 5{,}000-slot graylist, evicting $X$ head records removes an expected $X\cdot B/5{,}000$ benign entries.

\textit{\underline{Case A:} malicious replacement.}
When the selected peer is malicious, it samples records from the malicious peer pool established by \textit{N-I}.
This pool contains approximately $3{,}000{,}000$ distinct $\langle\mathrm{IP},\mathrm{port}\rangle$ pairs, roughly $600\times$ the graylist capacity.
The 500 returned records are therefore fresh with high probability and pass deduplication.
They FIFO-evict an expected $500\cdot B/5{,}000=B/10$ benign entries.
The expected net change is therefore $\Delta B=-B/10$.

\textit{\underline{Case B:} benign replacement.}
When the selected peer is benign, the target removes that benign record from the graylist, decreasing $B$ by one.
Under \textit{N-I}, the benign peer's whitelist is contaminated at approximately the $98.8\%$ occupation bound, so 500 returned records contain at most about six benign records.
The same 500 insertions FIFO-evict an expected $B/10$ benign entries.
The upper bound on the net change is therefore
\[
\Delta B\le -1+6-\frac{B}{10}=5-\frac{B}{10}.
\]
Thus Case B decreases $B$ when $B\ge50$ and can increase $B$ only slightly when $B<50$.

\section{Whitelist Attack Progression}
\label{app:nyx-whitelist-progression}

\begin{figure}[H]
    \centering
    \includegraphics[width=\linewidth]{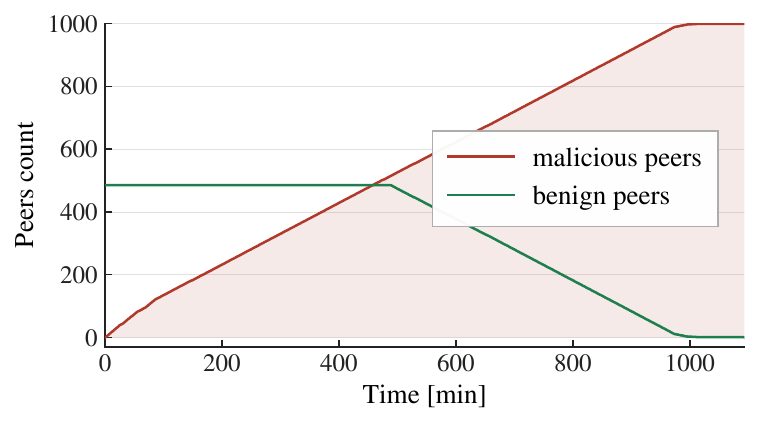}
    \caption{Progression of the whitelist attack on target node.}
    \label{fig:nyx_whitelist_long}
\end{figure}

\begin{figure}[H]
    \centering
    \includegraphics[width=\linewidth]{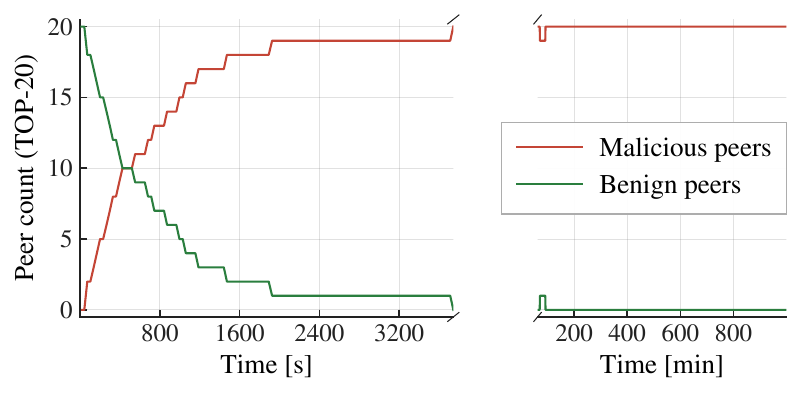}
    \caption{Status change of whitelist top-20 slots.}
    \label{fig:nyx_top20}
\end{figure}

This appendix extends \S\ref{subsec-eval-nyx-phase3} (whitelist
saturation paragraph) by showing the long-run dynamics of the
target's whitelist under \textsc{Nyx}.
Fig.~\ref{fig:nyx_whitelist_long} shows malicious and benign record
counts in the target's 1{,}000-slot whitelist over the entire
18.5\,h run.
The benign count starts at 483 (Table~\ref{tab:pre_attack_state}).
During the attack, both the Gray$\to$White promotion path and the
outgoing replacement path of \path|update_sync_search()| write only
malicious entries into the whitelist, so the benign count decreases
monotonically.
By $T_0+18.5$\,h, every whitelist slot is occupied by an
attacker-controlled record.
Fig.~\ref{fig:nyx_top20} zooms into the top-20 whitelist slots,
which have the highest \texttt{last\_seen} values and are therefore
selected first when the target picks outgoing peers.
The figure uses two time axes.
The seconds-scale axis on the left shows how benign top-20 entries
are quickly evicted during the first hour.
The minutes-scale axis on the right shows that the top-20 stays
fully attacker-controlled for the rest of the run.

\FloatBarrier

\section{Implementation Details (\S\ref{sec-eavlua})}
\label{app-impl-details}

Our implementation includes two major parts. 

\heading{Emulation environment}
We use the emulation environment to evaluate the full \textsc{Nyx} pipeline in a controlled setting. This allows us to run large-scale poisoning, simulate an unreachable target, and monitor peerlist and connection changes without affecting the public Monero network. We provide detailed configurations.

\textit{Multi-IP infrastructure.}
The attacker's 1{,}000 /24 subnets are provisioned as dummy IP interfaces on a dedicated router and advertised into the emulation fabric via BIRD. Whitelist poisoning runs from a controller that iterates through the 1{,}199 reachable nodes in batches of 120, launching 1{,}000 concurrent \textit{py-levin}~\cite{github-pylevin} instances per batch, each binding to one of the dummy IPs to issue a handshake request and respond to the \texttt{PING} probe for insertion. A listener module returns \texttt{PING} replies and terminates connections after insertion so that the attacker IPs remain available for subsequent batches. Graylist poisoning runs from 20 additional \textit{py-levin} containers that maintain persistent incoming connections with every reachable node and inject 250 distinct trash records per timed sync response.

\textit{Target NAT simulation.}
A host-side \texttt{iptables} rule drops all new inbound TCP connections to the target's P2P port while preserving \texttt{ESTABLISHED} return traffic, keeping the target's outbound connectivity intact.

\textit{Monitoring.}
A host-side RPC client queries each target node every 30\,s for the full whitelist, graylist, and outgoing connection set. Raw logs are stored per-node for post-experiment analysis.

\heading{Mainnet environment}
We use the mainnet environment to validate \textsc{Moros} under real deployment conditions while keeping the experiment narrowly scoped. The implementation focuses on the seed-node poisoning path and its effect on the bootstrapping behavior of a controlled newborn unreachable node.

\textit{Malicious peer pool.}
The attacker pre-constructs a pool of 6{,}000 distinct \texttt{<IP:Port>} records by assigning six P2P ports to each of its 1{,}000 public IP addresses, which span about eight /24 subnets. Each endpoint responds correctly to handshake and timed sync requests from the target. The pool size equals the combined capacity of the target's whitelist (1{,}000) and graylist (5{,}000).

\textit{Whitelist filling.}
The attacker deploys 1{,}000 virtual nodes against the six seed nodes. These nodes complete the handshake, respond to \texttt{PING} probes, and close the connection after insertion. They also maintain long-lived connections with the controlled target and return malicious peer records during timed sync.

\section{Reachable-Node Probing Infrastructure}
\label{app:probing}

The attacker maintains an updated inventory of Monero's reachable nodes to support the per-node dedicated-port assignment required by \textit{N-I} (\S\ref{sec-n1}).
The probing tool comprises two modules.

\heading{Node discovery.}
The discovery module scans candidate addresses that have not yet been confirmed online.
For each candidate, the module initiates a handshake to verify reachability and simultaneously collects the peerlist returned in the handshake response, expanding the candidate set for subsequent rounds.
By recursively crawling the peerlists of newly discovered nodes, the module progressively enumerates the reachable-node population without relying on a centralized directory.

\heading{Node confirmation.}
The confirmation module periodically re-checks nodes that have previously been confirmed reachable, assessing their long-term availability.
Nodes that fail confirmation across multiple consecutive rounds are marked offline and excluded from the filling target set.

\heading{Asynchronous probing.}
Both modules replace the stock Monero client's synchronous connection logic with asynchronous connection handling, enabling hundreds of concurrent handshake attempts from a single machine.
This modification reduces the time required to scan the full reachable-node population from hours to minutes.

\heading{Port assignment.}
Once the reachable-node inventory stabilizes, the attacker assigns each confirmed node a dedicated port number drawn from a pre-allocated range.
This mapping is stored as a lookup table indexed by the reachable node's \texttt{<IP:Port>} and is consulted during every whitelist-filling handshake to determine which port the attacker IP announces (\S\ref{sec-n1}, port-diversity design).

\heading{Scale.}
In our emulation environment, the probing tool discovers all 1{,}199 reachable nodes within the first round.
On the Monero mainnet, prior measurements report approximately 3{,}000 reachable nodes~\cite{cao2020exploring}, consistent with our probing results.

\section{Network Impairment Calibration}
\label{app:netem}

Our emulation runs all 1{,}200 Monero containers on a single physical server, yielding sub-millisecond inter-node latency (baseline RTT $< 1$\,ms).
To approximate real-world propagation, we inject network impairments via \texttt{tc/netem} on each container's network interface, with parameters calibrated against Monero mainnet measurements.

\heading{Mainnet measurement.}
From a server located in Europe, we probed 1{,}312 reachable Monero mainnet nodes with 20 ICMP pings per node and a TCP port-18080 connection test.
Table~\ref{tab:netem} summarizes the results.
98.9\% of nodes exhibited zero packet loss; the mean loss rate of 0.43\% is dominated by 15 nodes with severe loss.
The mean jitter (mdev) across all nodes is 1.38\,ms.

\begin{table}[t]
\centering
\caption{Monero mainnet latency measurements.}
\label{tab:netem}
\begin{tabular}{lrrrrr}
\toprule
\multicolumn{1}{c}{\textbf{Metric}} & \textbf{Min} & \textbf{Median} & \textbf{Mean} & \textbf{P95} & \textbf{Std} \\
\cmidrule{2-6}
ICMP RTT (ms) & 7 & 100 & 117.3 & 247.9 & 70.9 \\
TCP connect (ms) & 7.3 & 101 & 121.1 & 253.7 & --- \\
\bottomrule
\end{tabular}
\end{table}

\heading{Impairment configuration.}
Based on the measurements above, each container receives the following \texttt{netem} parameters: 50\,ms one-way delay (yielding an end-to-end RTT of ${\sim}$100\,ms, matching the measured median), 1.5\,ms jitter, 0.1\% packet loss, and a bandwidth cap of 100\,Mbit/s.
Parameters are randomized at the AS level (delay $\pm 20\%$, loss $\pm 30\%$) with a fixed random seed for reproducibility.

\heading{Validation.}
After injection, we sampled cross-AS node pairs and confirmed a median RTT of ${\sim}$100\,ms with near-zero packet loss, consistent with the mainnet conditions reported above.

\section{The 6-Group Strategy and Its Constant-Cost Property}
\label{app:sixgroup}

The port-diversity design in \S3.2.1 does not require a dedicated port per
reachable node; it can be realized with a fixed set of ports. We instantiate
it as the \emph{6-group} strategy.

\paragraph{Adversarial setup.}
The adversary provisions $6{,}000$ node-identities from $1{,}000$ physical IPs,
each hosting six listening ports, and divides them into six groups by port.
This budget exactly covers the combined capacity of a target's whitelist
($1{,}000$) and graylist ($5{,}000$), i.e., $6{,}000$, which suffices to fill
the target's peerlist.

\paragraph{Grouped filling with port rotation.}
During filling, the adversary randomly partitions all reachable nodes into six
groups, each poisoned by the corresponding group of ports. To ensure that all
$6{,}000$ records are accepted by the target without being merged under
$\langle \text{IP}, \text{port}\rangle$ deduplication, the adversary does not
fix the partition but periodically rotates the port assignment of each
reachable node. With rotation, even if several of the target's outgoing peers
fall into the same group, the records carried toward the target remain mutually
distinct in the $\langle \text{IP}, \text{port}\rangle$ space and continue to
pass the target's graylist deduplication (\S2).

\paragraph{Evaluation.}
We validate the 6-group strategy under the same experimental environment as
\S4.1. Over an approximately 20-hour run, with the adversarial identities
reduced to $6{,}000$, the attack still achieves a complete eclipse, reaching
$\mathrm{CTR}=100\%$ within about 30 minutes, consistent with the N-group
evaluation in \S4.3, even though the number of adversarial identities drops
sharply from $1{,}000\times1{,}200$ to $1{,}000\times6$.

\paragraph{Distinction from Sybil flooding.}
The defining property of this instantiation is constant cost, i.e., the
adversarial budget is fixed at $6{,}000$
$\langle \text{IP}, \text{port}\rangle$ identities. Since the peerlist capacity
and the poisoning required per target do not grow with the network size, this
fixed budget suffices, through Monero's peerlist management mechanism (\S2), to
keep each target's peerlist dominated by adversarial records. This is the
essential distinction from trivial Sybil flooding, which requires the adversary
to hold an overwhelming majority of nodes relative to the whole network, with
cost growing with the network size, whereas our attack only needs a fixed
number of identities far smaller than the benign population.

\section{Measuring the Unreachable Fraction}
\label{app:unreach}

We measured the unreachable fraction on the mainnet. Four geographically
diverse vantage points (DE/UK/Beijing/US) observed all incoming connections
over ${\sim}70$\,h ($35{,}489$ handshakes); each connecting node is classified
by back-pinging the listening port it declares in its handshake: a pong means
reachable, otherwise unreachable.

Experimental results show that $80.64\%$ of nodes are unreachable
($12{,}977$ of $16{,}092$ unique \texttt{peer\_id}), stable across all four
vantage points ($79.21$--$82.05\%$). Independent RPC snapshots of live incoming
connections (every $10$\,min) match the handshake log for $95.95\%$ of sampled
connections, confirming that the log captures essentially all observed peers.

\end{document}